\documentclass[notoc,paper,letterpaper]{JHEP3}
\usepackage{amssymb}
\usepackage{epsfig}
\newcommand{\be}{\begin{equation}}
\newcommand{\ee}{\end{equation}}
\newcommand{\bea}{\begin{eqnarray}}
\newcommand{\eea}{\end{eqnarray}}
\newcommand{\ie}{{\it i.e.}}
\newcommand{\eg}{{\it e.g.}}
\newcommand{\etc}{{\it etc.}}

\newcommand{\U}{\mathop{\rm U}}
\newcommand{\SU}{\mathop{\rm SU}}

\newcommand{\Sp}{\mathop{\rm Sp}}
\newcommand{\bo}{\hbox{1\kern-.23em{\rm l}}}
\newcommand{\til}{\widetilde}
\def\del{{\partial}}
\def\delb{{\bar\del}}
\def\bar{\overline}

\def\a{\alpha}
\def\b{\beta}

\def\de{\delta}
\def\ad{{\dot\a}}
\def\bd{{\dot\b}}

\def\ep{\epsilon}
\def\si{\sigma}
\def\th{\theta}
\def\thb{{\bar\th}}
\def\thp{\th^+}
\def\thm{\th^-}
\def\thbp{\thb^+}
\def\thbm{\thb^-}
\def\psib{{\bar\psi}}
\def\abar{{\bar a}}

\def\Ibar{{\bar I}}
\def\Jbar{{\bar J}}

\def\Lbar{{\bar L}}
\def\Wb{{\bar W}}
\def\qt{{\tilde q}}
\def\qp{q^+}
\def\qm{q^-}
\def\qtp{{\qt^+}}
\def\qtm{{\qt^-}}
\def\Qb{{\bar Q}}
\def\Db{{\bar D}}
\def\Dp{D^+}
\def\Dm{D^-}
\def\Dbp{\Db^+}
\def\Dbm{\Db^-}
\def\Dpp{{D^{++}}}
\def\Dmm{{D^{--}}}
\def\Vpp{{V^{++}}}
\def\Vmm{{V^{--}}}
\def\up{u^+}
\def\um{u^-}
\def\fb{{\bar f}}
\def\fp{f^+}
\def\fm{f^-}
\def\fbp{\fb^+}
\def\fbm{\fb^-}
\def\dsl{{\not\!\del}}
\def\dpp{{\not\!\del^{++}}}
\def\dpm{{\not\!\del^{+-}}}
\def\dmp{{\not\!\del^{-+}}}
\def\dmm{{\not\!\del^{--}}}

\title{Higher-Derivative Terms in N=2 Supersymmetric Effective Actions}

\author{Philip C. Argyres, Adel M. Awad, Gregory A. Braun, and F. Paul Esposito\\
Department of Physics, University of Cincinnati, Cincinnati OH 45221-0011 \\ 
E-mail: \email{argyres,adel,braun,esposito@physics.uc.edu} }

\abstract{We show how to systematically construct higher-derivative
terms in effective actions in harmonic superspace despite the infinite
redundancy in their description due to the infinite number of
auxiliary fields.  Making an assumption about the absence of certain
superspace Chern-Simons-like terms involving vector multiplets, we write 
all 3- and 4-derivative terms on Higgs, Coulomb, and mixed branches.  Among 
these terms are several with only holomorphic dependence on fields, and at
least one satisfies a non-renormalization theorem.  These holomorphic terms 
include a novel 3-derivative term on mixed branches given as an integral
over 3/4 of superspace.  As an illustration of our method, we search for 
Wess-Zumino terms in the low energy effective action of $N=2$ supersymmetric 
QCD.  We show that such terms occur only on mixed branches.  We also present
an argument showing that the combination of space-time locality with supersymmetry
implies locality in the anticommuting superspace coordinates of for unconstrained
superfields.}

\preprint{UCTP-106-03}

\begin{document}

\section{Introduction and auxiliary fields in derivative expansions}

Certain higher-derivative terms in the effective actions of four
dimensional gauge theories with extended supersymmetry have been shown
to be not renormalized \cite{ds9705,dg9909}.  Another class of
non-renormalized higher derivative terms are the Wess-Zumino (WZ)
terms, shown to exist, though their fully supersymmetric form
was not determined, by a one-loop calculation in \cite{tz9911} and by
an anomaly matching argument \cite{i0001} for $N=4$ supersymmetric
effective actions.  They must therefore also exist in $N=2$ effective
actions.  In this paper we will derive the fully supersymmetric form
of these $N=2$ supersymmetric WZ terms, (they were already found for
$N=1$ supersymmetric effective actions in \cite{nr85}).  In doing so,
we will develop the tools to perform a systematic exploration of 
higher-derivative terms in $N=2$ effective actions, and will carry
out this exploration to construct all terms up to and including
4 derivatives, many of which are constrained by new non-renormalization 
theorems.

Although higher-derivative terms in the low energy effective actions of four
dimensional gauge theories with extended supersymmetry have received
some attention \cite{h9507,dgr9601,glrv9607,crv9612,gr9804,bk9804,gkpr9810,
bbk9810,lv9811,bkt9911,bbp0008,km0101,km0105,bi0111,bbp0205,bip0210,bbp0304},
no systematic exploration of the $N=2$ case has been done.  This may be due to the
difficulty of generating higher-derivative terms with extended
supersymmetry.  In an on-shell and/or component formalism, the problem
is that one must self-consistently correct the supersymmetry
transformation rules order by order in the derivative expansion at the
same time that one tries to construct the supersymmetry-invariant
higher-order term in the action.  The solution to this problem is to
use an off-shell superfield formulation so that the supersymmetry
transformations are independent of the form of the action.  In this
case, it only remains to list all the supersymmetry invariants with a
given number of derivatives.  A prescription for generating all
possible such terms might only exist if the superfields are
unconstrained; the constrained case is unclear.  
Harmonic superspace \cite{gikos84} gives such an
unconstrained superfield formulation for $N=2$ supersymmetry.  We
therefore use the harmonic superspace formalism in this paper.  An
important feature of harmonic superspace is that, in addition to the
usual space-time directions described by coordinates $x^\mu$ and
Grassmann-odd directions with spinor coordinates $\th^i_\a$, there is
also a 2-sphere described by commuting harmonic coordinates
$u^\pm$; see, \eg, \cite{gios01} or section 3 below for a review of
harmonic superspace.

The low energy effective action at a generic vacuum of $N=2$ gauge
theory includes only massless $\U(1)$ vector multiplets and massless
neutral hypermultiplets, since charged hypermultiplets generically get
masses by the Higgs mechanism.  We call the set of those vacua with only
massless neutral hypermultiplets the ``Higgs branches", those with only
$\U(1)$ vector multiplets the ``Coulomb branch", and those vacua with
both kinds of multiplets the ``mixed branches" \cite{sw9407,sw9408,aps9603}.
Thus the low energy propagating fields are massless neutral scalars
$\phi$ and spinors $\psi_\a$, and $\U(1)$ vectors $A^\mu$.

We organize the terms in a low energy effective action
by a kind of scaling dimension which essentially counts the number of
derivatives, and we refer to this counting as the ``dimension'' and
denote it by square brackets.  It should be noted that it is not
the same as the scaling dimension used in the renormalization group
analysis of fluctuations around a given vacuum.  The leading term\footnote{We
are considering here theories without Fayet-Iliopoulos terms, \eg\ any
$N=2$ gauge theory whose microscopic description involves only semi-simple
gauge groups.} in the effective Lagrangian for the scalar fields is, 
schematically,
\be\label{2dea}
g(\phi) \del^\mu\phi\del_\mu\phi .
\ee
Since this is a 2-derivative term we assign it
dimension 2, so we must assign $[\del_\mu]=1$ and $[\phi]=0$.
Supersymmetry then determines the dimensions of the other fields: the
supersymmetry algebra implies that $-[\th] = [d\th] = [\del/\del\th] =
1/2$, the normalization of the harmonic sphere $u^\pm$ coordinates (\eg,
eqn.\ \ref{uident}\ below) implies that $[u] = [\del/\del u] = [du] =
0$, and examination of the component expansion of hypermultiplet and
vector multiplet superfields then implies that $[\psi_\a] = 1/2$ and
$[A_\mu] = 0$.

A drawback of the harmonic superspace formalism, or any unconstrained
off-shell superfield formalism describing hypermultiplets, is that its
superfields include an infinite number of auxiliary component fields.
This implies that at each order of the derivative expansion of an
effective action, we are able to write an infinite number of terms of
that order in harmonic superspace, even though, when written in terms
of propagating component fields (\ie, after substituting for the
auxiliary fields by their equations of motion), there is only a
finite number of such terms.\footnote{Note that we do not count the
infinite number of coefficient functions of the dimensionless
propagating scalars, such as $g(\phi)$ in (\ref{2dea}) as separate
terms.}  This infinite redundancy in the harmonic
superfield formalism appears because the Lagrangian can be 
non-local with respect to the harmonic 2-sphere coordinates, or,
equivalently, because terms with arbitrarily many $\del/\del u$
derivatives enter at each order of the derivative expansion.
There is no physical requirement of locality in the
auxiliary $u$ variables.

(The same argument could be made for the auxiliary Grassmann variables
$\theta$ in superspace: Since there is no physical requirement that
effective actions be local in the Grassmann coordinates, doesn't this
mean that every superspace description, not just harmonic superspace,
suffers from this kind of redundancy?  It turns out
that the combination of locality in space-time together with supersymmetry
invariance and the anticommuting nature of Grassmann coordinates implies
that any term in the effective action of unconstrained superfields can
be written as the integral of a local functional in the Grassmann coordinates.
This argument is presented in appendix A below, and will be useful in the
discussion of $N=2$ vector multiplets in section 5.)

A worry is then that this redundancy following from non-locality in the
$u^\pm$ coordinates makes the harmonic superspace
formalism useless for systematic derivative expansions of effective
actions, for there is no simple way of listing or parameterizing terms
with arbitrary $\del/\del u$ derivative dependence.  General
considerations involving the nature of auxiliary fields and of
derivative expansions, however, imply that there is a finite,
order-by-order, procedure for constructing all terms in the effective
action given the leading term, as we will now show.

Consider a theory with some propagating fields and some auxiliary
fields which we collectively denote by $p$ and $a$, respectively.
Suppose we are given some leading (2-derivative) action $S_2(p,a)$,
such that the $\del_a S_2 = 0$ equations of motion determine the
auxiliary fields in terms of the propagating fields and their
derivatives,
\be\label{a2eom}
a = a_2(\del,p) ,
\ee
as is the case with the 2-derivative hypermultiplet and vector
multiplet actions in harmonic superspace.  Thus, $a_2$ are the
functions satisfying
\be\label{ds2ata2}
\del_a S_2 |_{a = a_2} = 0 .
\ee
Now the general effective action is a sum of contributions $S_n$ with
$n\ge2$ derivatives:
\be\label{fullact}
S = S_2 + \ell S_3 + \ell^2 S_4 + \cdots ,
\ee
where $\ell$ is the cut-off length scale which organizes the
derivative expansion.  In an effective action
expansion, we develop the fields in a power series expansion in $\ell$.
In particular, the solution to the equations of motion for the
auxiliary fields,
\be\label{fulleom}
0 = \del_a S = \del_a S_2 + \ell \del_a S_3
+ \ell^2 \del_a S_4 + \cdots ,
\ee
following from (\ref{fullact}), will be the leading piece (\ref{a2eom})
plus corrections of order $\ell$:
\be\label{aexp}
a = a_2(\del,p) + \ell a_3(\del,p) + \ell^2 a_4(\del,p) + \cdots .
\ee
Plugging (\ref{aexp}) back into (\ref{fulleom}) and expanding in
powers of $\ell$, we then determine $a_n$ in terms of the $a_m$, $m<n$:
\bea\label{aneom}
a_3 &=& - \left(S_2''|_2\right)^{-1} S_3'|_2 ,\nonumber\\
a_4 &=& - \left(S_2''|_2\right)^{-1} \left( S_4'|_2 +
a_3 S_3''|_2 + \textstyle{1\over2} a_3^2 S_2'''|_2 \right)
,\nonumber\\ && \cdots ,
\eea
where primes denote derivatives with respect to $a$ and $|_2$ means
evaluate at $a=a_2$.  Substituting (\ref{aexp}) back into the full
action (to eliminate the auxiliary fields) and expanding in powers of
$\ell^2$, we have
\bea\label{onshellact}
S &=& S_2|_2 + \ell \left[ a_3 S_2'|_2 + S_3|_2 \right]
+ \ell^2 \left[ a_4 S_2'|_2 + \textstyle{1\over2} a_3^2 S_2''|_2
+ a_3 S_3'|_2 + S_4|_2 \right] + \cdots \nonumber\\
&=& S_2|_2 + \ell S_3|_2
+ \ell^2 \left[ S_4 - \textstyle{1\over2} S_3' (S_2'')^{-1} S_3'
\right]|_2 + \cdots ,
\eea
where in the second line we have used (\ref{ds2ata2}) and
(\ref{aneom}).  The crucial point is that when the auxiliary fields
are eliminated by substitution, 
the $n$-derivative piece of the action, appearing
at order $\ell^{n-2}$, depends on $S_n$ only through $S_n|_2$, \ie,
with the auxiliary fields evaluated at their values $a=a_2$ determined
from the leading ($S_2$) term in the derivative expansion.  This means
that in classifying the $S_n$ terms, {\em it is sufficient to
substitute in the auxiliary component fields in terms of the
propagating components using (\ref{a2eom}).}  Since there are only a
finite number of terms in $S_n$ when written in terms of the
propagating components, it follows that only a finite number of
superfield expressions need be examined at this order, despite their
infinite number of possible forms.  In particular, this means that
there will be identities relating the superfields evaluated at $a=a_2$
to some number of their $\del/\del u$ derivatives which can be used to
truncate the $\del/\del u$ expansion at each order in the space-time
derivative expansion.  Explicit examples of this procedure appear in
sections 4 and 6 below.


The above argument is not enough to assure us that we can use the
harmonic superspace techniques to perform systematic derivative
expansions.  Two further problems may arise.  First, the above
argument assumed that the leading, 2-derivative, term was already
given, and did not tell us how to remove the infinite redundancy in
its harmonic superspace description.  Fortunately this hard work has
already been done for us (see, \eg, Chapters 7 and 11 of
\cite{gios01}): superspace expressions for all $N=2$ supersymmetric
2-derivative terms for $\U(1)$ vector multiplets and neutral
hypermultiplets have been found.

A second potential problem is that a systematic derivative expansion
can only be carried out if the superfields all have non-negative
dimension.  For suppose that a field had negative dimension: then a
term of given overall dimension may have an arbitrarily large number
of positive-dimension derivatives as well as negative-dimension
fields.  This problem does not arise for the hypermultiplet superfield
$q^+$ since it turns out to have dimension $0$.  The gauge invariant
field strength superfield $W$ for the vector multiplets also has
dimension $0$.  However, it is a constrained superfield as it must
satisfy the Bianchi identities.  The unconstrained superfield is the
vector potential superfield $\Vpp$ which has dimension $-1$.  Actions
built from the $\Vpp$ must be gauge invariant.  This acts as a
powerful restriction on the ways in which the $\Vpp$ can enter.  In
particular, the question arises whether there can exist Chern-Simons-like
terms in $N=2$ harmonic superspace, \ie, do there exist gauge
invariant $\U(1)$ vector multiplet terms which cannot be written solely
in terms of the field strength multiplets $W$?  If not, then we can
just work with the dimension $0$ field strength superfield $W$.  The
existence of superspace Chern-Simons-like terms is a difficult algebraic
question; they are known to
occur, for example, in $N=3$ harmonic superspace \cite{r83,gikos85,gio87,gios01}.
In section 5 below we show that in $N=2$ harmonic superspace any gauge-invariant
term written in terms of the potential superfield $\Vpp$ can be rewritten
solely in terms of derivatives of the field strength superfield $W$,
at the expense of introducing non-localities in the Grassmann coordinates.
The argument of appendix A, showing locality in the Grassmann directions of
superspace, fails when the superfields obey extra constraints, in this case 
the Bianchi identities.  
\emph{Thus we show that the existence of superspace Chern-Simons-like
terms is equivalent to the existence of supersymmetric expressions involving
field-strength superfields non-local in the Grassmann coordinates.}
Due to the nilpotent nature of Grassmann variables, the number of
such possible non-local terms is finite at each order in the derivative
expansion.  We will explore the possibility of superspace Chern-Simons-like
terms with three or four derivatives elsewhere \cite{aabe0307}, and, for 
the purposes of this paper, we will assume they do not occur.

\section{Summary of results}

Modulo the question of the existence of superspace Chern-Simons-like terms,
we find the following possible harmonic superspace forms for 3- and 
4-derivative terms in the low energy effective action of $N=2$ supersymmetric
theories.  On the Higgs branches, with only neutral hypermultiplets $\qp_I$ and
their complex conjugates $\qm_\Ibar$,
there are no 3-derivative terms, and two types of 4-derivative terms (harmonic
superspace notation is reviewed in section 3 below):
\bea\label{higgsres}
S^H_{4a} &=&  \int\! du\, d^4x\,\, 
d^2\thp\, d^2\thbp\,\,
\del^\mu \qp_I \del_\mu \qp_J\  
B^{IJ}(\qp_K; u^\pm, \Dpp) +\mbox{c.c.},\nonumber\\
S^H_{4b} &=&  \int\! du\, d^4x\,\, 
d^2\thp\, d^2\thbp\, d^2\thm\, d^2\thbm\,\,
\Gamma(\qp_I,\qm_\Ibar; u^\pm, D^{\pm\pm}) .
\eea
Similarly, on the Coulomb branch, with only $\U(1)$ field strength vector 
multiplets $W_a$ and their complex conjugates $\Wb_\abar$, we find
\bea\label{coulres}
S^C_{4a} &=& \int\! d^4x\,\, 
d^2\thp\, d^2\thm \,\,
\del^\mu W_a \del_\mu W_b\  {\cal G}^{ab}(W_c) +\mbox{c.c.},\nonumber\\
S^C_{4b} &=& \int\! d^4x\,\, 
d^2\thp\, d^2\thbp\, d^2\thm\, d^2\thbm\,\,
{\cal H}(W_a,\Wb_\abar) .
\eea
On the mixed branches there are both 3-derivative and 4-derivative terms:
\bea\label{mixedres}
S^M_3 &=& \int\! du\, d^4x\,\,
d^2\thp\, d^2\thbp\, d^2\thm\,\,
F(\qp_I,W_a; u^\pm,\Dpp) +\mbox{c.c.},\nonumber\\
S^M_{4a} &=& \int\!du\, d^4x\, d^2\thp\, d^2\thbp\, d^2\thm 
\,\,\Dp W_a\cdot\Dp W_b\ G^{ab}(\qp_I, W_c; u^\pm, \Dpp) + \mbox{c.c.},
\nonumber\\
S^M_{4b} &=& \int\!du\, d^4x\, d^2\thp\, d^2\thbp\, d^2\thm 
\,\,\Dbm(\Dpp)^n\qp_I\cdot \Dbm(\Dpp)^m\qp_J \  
G_{nm}^{IJ}(\qp_K, W_a; u^\pm, \Dpp) 
+ \mbox{c.c.}, \nonumber\\
S^M_{4c} &=& \int\! du\, d^4x\,\,
d^2\thp\, d^2\thbp\, d^2\thm\, d^2\thbm\,\,
H(\qp_I,\qm_\Ibar,W_a,\Wb_\abar; u^\pm,D^{\pm\pm}) .
\eea
For each of the mixed branch terms given as integrals over 3/4
of superspace ($S^M_3$, $S^M_{4a}$, and $S^M_{4b}$), there is another
term given by an integral over a different three-quarters of superspace,
for example:
\be
S^{'M}_3 = \int\! du\, d^4x\,\,
d^2\thp\, d^2\thbp\, d^2\thbm\,\,
F'(\qp_I,\Wb_\abar; u^\pm,\Dpp) +\mbox{c.c.}.
\ee

The terms $S^H_{4a}$, $S^C_{4a}$, $S^M_3$, and $S^M_{4a,b}$ do 
not seem to have been noted elsewhere in the literature.  They
depend only on the analytic hypermultiplets and chiral vector 
multiplets and not their complex conjugates.  The holomorphic 
nature of these terms suggests that they might be determined 
non-perturbatively using arguments similar to those of \cite{sw9407,sw9408}.

Indeed, in $N=2$ superQCD, where the strong coupling scale $\Lambda$ can
be thought of as the lowest component of a field strength vector superfield
$W$, $S^H_{4a}$ can get no $\Lambda$-dependent quantum corrections since it 
cannot involve any $W$'s.  Thus $S^H_{4a}$ satisfies a non-renormalization 
theorem.  (On the other hand, $S^H_{4b}$ {\em can} get quantum corrections 
because we can add $W$-dependence to it as in $S^M_{4c}$.  In this sense, 
for the puposes of deriving non-renomalization theorems, we should think of 
$S^H_{4b}$ and $S^C_{4b}$ as special cases of $S^M_{4c}$.)
 
Similarly, the Coulomb branch holomorphic 4-derivative term $S^C_{4a}$ can
only get quantum corrections holomorphic in $\Lambda$, \ie, only one loop and
instanton corrections.  Note that when there is only a single vector multiplet,
$S^C_{4a}$ can be rewritten using the Bianchi identity as an $S^C_{4b}$ term; 
see the discussion after (\ref{CBterms}) below.  Thus examples of $S^H_{4a}$ 
terms only occur with two or more vector multiplets.

Finally, the holomorphic 3-derivative terms on the mixed branch, $S^M_3$ and 
$S^{'M}_3$, are of special interest since they also only get one loop and
instanton corrections, and they give the entire leading correction to the 
mixed branch low-energy physics.  They describe a derivative coupling between
the hypermultplet scalars and the vectormultiplet photons; see (\ref{SM3comp}) 
below.

The expressions in (\ref{higgsres}) and (\ref{mixedres}) are non-local in 
the $u^\pm$ variables, since they involve infinitely many $D^{\pm\pm}$ 
derivatives in gerneral.  However, we will show that only a certain finite number of 
combinations of those derivatives may act on any given $q^\pm$ field in 
these expressions.  For example, we show below that if the leading 
2-derivative term for the hypermultiplets describes free hypermultiplets, 
then only the combinations $\qp$, $\Dpp\qp$, and $(\Dpp)^2\qp$ may appear 
in $S^H_{4a}$, $S^M_3$, $S^M_{4a}$, and $S^M_{4b}$; while the 
non-holomorphic terms $S^H_{4b}$ and $S^M_{4c}$ can be taken 
to depend only on the combinations $(\Dmm)^3\qp$, 
$(\Dmm)^2\qp$, $\Dmm\qp$, $\qp$, $\Dpp\qp$, and $(\Dpp)^2\qp$, and their 
complex conjugates involving $\qm$.  Thus, in terms of these sets of 
fields, the effective actions are local in the harmonic superspace variables 
$u^\pm$ as well as the $x^\mu$.  For a more general 2-derivative term, the 
description of this finite set of hypermultiplet fields is more complicated, 
though it can be derived in principle.  Furthermore, using this 
characterization of 4-derivative terms we show that WZ terms can only 
occur on mixed branches in $N=2$ effective actions.

The outline of the rest of this paper is as follows.  In section 3 we briefly 
review the harmonic superspace formalism.  In section 4 we characterize the 
4-derivative terms made out of hypermultiplets and we show that no WZ terms 
can be constructed purely from hypermultiplets.  In section 5 we do the same 
for $\U(1)$ field strength vector multiplets, where we also discuss the problem
of superspace Chern-Simons-like terms.  Finally, in section 6 we characterize 
the 3- and 4-derivative terms with both vector and hypermultiplets, and 
construct WZ terms.

\section{Harmonic superspace and notation}

We briefly summarize harmonic superspace formalism following the
notation and conventions of \cite{gios01}.  $N=2$ supersymmetry
without central charges has two fermionic generators, $Q^i$, $i=1,2$,
satisfying $\{Q_\a^i,\bar Q_{\ad j}\} = 2 \de^i_j\si^\mu_{\a\ad} P_\mu$,
with the other anticommutators vanishing.  Since we need only consider
neutral fields in the low energy effective action, there are no
central charges.
Harmonic superspace allows an unconstrained superfield formulation
of $N=2$ supersymmetry by permitting an infinite number of auxiliary
fields.  This is done in a superspace consisting of a standard
superspace with coordinates $\{x^\mu,\th^i_\a,\thb_{\ad i}\}$ extended
by two even coordinates on an additional 2-sphere.  The $\th$'s are
Grassmann spinor coordinates satisfying the complex conjugation rule
\be\label{thcc}
\bar{\th^i_\a} = - \thb_{\ad i} .
\ee

The additional 2-sphere is conveniently coordinatized by introducing
harmonic $\SU(2)$ group coordinates $u^\pm_i$.  Here $i$ is an
$\SU(2)$ index, while the $\pm$ indices refer to the diagonal
$\U(1)\subset\SU(2)$ charge.  The $u^\pm$ variables satisfy the
following basic identity,
\be\label{uident}
\up_i \um_j - \up_j \um_i = \ep_{ij} ,
\ee
along with the complex conjugation rule
\be\label{ucc}
\bar{u_i^\pm} = \mp u^{i\mp}.
\ee
Restriction to the coset sphere $S^2 = \SU(2)/\U(1)$ is realized by
having all physical expressions be $\U(1)$ neutral.  (Note that the
$\SU(2)$ indices $i,j,k,\ldots$ are raised and lowered with the
anti-symmetric $\ep^{ij}$ tensor defined by $\ep_{12}=-\ep^{12}=1$, so
that $a^i = \ep^{ij} a_j$, $a_i = \ep_{ij} a^j$.)  Thus the full
harmonic superspace consists of the space-time, Grassmann, and harmonic
coordinates $\{x^\mu,\th^i_\a,\thb_{\ad i},u^\pm_i\}$.

A basic assumption of the harmonic superspace formalism is that all
fields are harmonic functions on the sphere, which is to say they are
given by a power series expansion in the $u^\pm_i$ coordinates.  Due
to the identity (\ref{uident}), any product of $u^\pm_i$'s can be
rewritten as a sum of terms each completely symmetric on $\SU(2)$
indices.  For example, the expansion for a field of $\U(1)$ charge +1
will have the unique expansion $f^{+} = f^i \up_i + f^{(ijk)}
\up_i\up_j\um_k +\cdots$.

The usual superspace covariant derivatives are introduced
\be\label{covder1}
D^i_\a \equiv {\del\over\del\th^\a_i}+i\thb^{\ad i} \dsl_{\a\ad},
\qquad
\Db_{\ad i} \equiv -{\del\over\del\thb^{\ad i}} - i\th^\a_i \dsl_{\a\ad},
\ee
where $\dsl_{\a\ad}= \si^\mu_{\a\ad}\del_\mu$, satisfying the $N=2$
algebra $\{D^i_\a,\Db_{\ad j}\}=-2i\de^i_j \dsl_{\a\ad}$.  On the
sphere we introduce derivatives
\be\label{Ddef}
\Dpp \equiv \up_i {\del\over\del\um_i} , \qquad
\Dmm \equiv \um_i {\del\over\del\up_i} , \qquad
D^0 \equiv  \up_i {\del\over\del\up_i} - \um_i {\del\over\del\um_i} ,
\ee
which satisfy the $\SU(2)$ algebra
\be\label{su2alg}
[D^0,D^{\pm\pm}] = \pm 2D^{\pm\pm}, \qquad [\Dpp,\Dmm] = D^0.
\ee
Likewise, the usual space-time and Grassmann integration measures are
introduced, as well as a measure $du$ for integration over the sphere
satisfying
\be
\int\! du = 1 , \qquad
\int\! du\, f^{(q)}(u) = 0 \ \ \mbox{if}\ q \neq 0 , \qquad
\int\! du \, \up_{(i_1} ... \up_{i_n}\um_{j_1} ... \um_{j_n)} = 0,
\ee
where $f^{(q)}$ is any field of $\U(1)$ charge $q$.

It is useful to introduce the harmonic-projected Grassmann variables
\be\label{thpmdef}
\th^\pm_\a \equiv u^\pm_i \th^i_\a , \qquad
\thb^\pm_\ad \equiv u^\pm_i \thb^i_\ad ,
\ee
their derivatives,
\be
\del^\pm_\a \equiv {\del\over\del\th^{\mp\a}}
= \pm u^\pm_i {\del\over\del\th^\a_i}, \qquad
{\bar\del}^\pm_\ad \equiv {\del\over\del\thb^{\mp\ad}}
= \pm u^{\pm i} {\del\over\del\thb^{\ad i}},
\ee
and the associated harmonic-projected covariant derivatives
\bea\label{Dpmdef}
D^\pm_\a &\equiv& \phantom{-} u^\pm_i D^i_\a
= \pm\del^\pm_\a +i \thb^{\pm\ad}\dsl_{\a\ad} , \nonumber\\
\Db^\pm_\ad &\equiv& - u^{\pm i} \Db_{\ad i}
= \pm{\bar\del}^\pm_\ad -i \th^{\pm\a}\dsl_{\a\ad} .
\eea
We use the set $\{x^\mu,\th^\pm_\a,\thb^\pm_\ad,u^\pm_i\}$ as a
coordinate basis---called the {\em central basis}\footnote{Note that
this is slightly different from the meaning of central basis in
\cite{gios01} who use the $\th^i$'s instead of the $\th^\pm$'s.}---for
harmonic superspace.  Notice that in changing basis from the $\th^i$'s
to the $\th^\pm$'s the harmonic derivatives (\ref{Ddef}) pick up extra
terms, \eg, $\Dpp = \up_i (\del/\del\um_i) + \th^{+\a}\del^+_\a +
\thb^{+\ad}{\bar\del}^+_\ad$, \etc.  The harmonic covariant derivatives
then obey together with (\ref{su2alg}) the algebra
\bea\label{Dalg}
{}[D^{\pm\pm},D^\mp_\a] = D^\pm_\a, \ &\ & \
{}[D^{\pm\pm},\Db^\mp_\ad] = \Db^\pm_\ad, \nonumber\\
\{D^\pm_\a,\Db^\mp_\ad\} &=& \mp 2i\dsl_{\a\ad} ,
\eea
with all other (anti)commutators vanishing.  Eqs.\ (\ref{Dalg}) and
(\ref{su2alg}) give the form of the $N=2$ algebra on harmonic
superspace that we will use.

$N=2$ supersymmetry invariants can be formed by integrating a general
harmonic superfield over all the superspace coordinates with measure
$\int du\, d^4x\, d^4\thp\, d^4\thm$, where, up to total space-time
derivatives,
\be\label{defDpm4}
\int\! d^4\th^\mp = (D^\pm)^4
\equiv {1\over16} (D^\pm)^2 (\Db^\pm)^2 ,
\ee
where the derivatives are evaluated at $\th^\mp = 0$.

Two different constraints in $N=2$ harmonic superspace can
be used to reduce superfield representations.  We refer to these
two conditions as the {\em chiral constraint} and the (Grassmann)
{\em analytic constraint}, respectively.

The chiral constraint on a general superfield $\Phi$,
\be\label{CC}
\Dbp_\ad \Phi = \Dbm_\ad \Phi = 0,
\ee
is consistent since $\{\Dbp_\ad,\Dbm_\bd\}=0$, and can be solved
by introducing the chiral space-time coordinate
\be\label{xc}
x_C^\mu \equiv x^\mu - i \thp\si^\mu\thbm + i \thm\si^\mu\thbp
\ee
annihilated by $\Db^\pm$.  Then, in the {\em chiral basis}
$\{x_C^\mu,\th^\pm_\a,\thb^\pm_\ad,u^\pm_i\}$, the chiral constraint
can be solved by an arbitrary (unconstrained) superfield independent
of the $\thb^\pm$'s: $\Phi = \Phi(x_C^\mu,\th^\pm_\a,u^\pm_i)$.  These
{\em chiral superfields} are useful for describing the field-strength
superfield for the vector multiplet.  Supersymmetry invariants
can be constructed by integrating chiral superfields against the measure
\be\label{cmeas}
\int\! du\, d^4x_C\, d^4\th = \int\! du\, d^4x\, D^4 ,
\ee
where
\be\label{defD4}
D^4 \equiv {1\over16} (\Dp)^2 (\Dm)^2 .
\ee

The analytic constraint on a general superfield $\Phi$,
\be\label{AC}
\Dp_\a \Phi = \Dbp_\ad \Phi = 0,
\ee
is consistent since $\{\Dp_\a,\Dbp_\ad\}=0$, and can be solved
by introducing the analytic space-time coordinate
\be\label{xa}
x_A^\mu \equiv x^\mu - i \thp\si^\mu\thbm - i \thm\si^\mu\thbp
\ee
annihilated by $\Dp$ and $\Dbp$.  Then, in the {\em analytic basis}
$\{x_A^\mu,\th^\pm_\a,\thb^\pm_\ad,u^\pm_i\}$, the analytic constraint
can be solved by an arbitrary (unconstrained) 
superfield independent of $\thm$ and
$\thbm$: $\Phi = \Phi(x_A^\mu,\thp_\a,\thbp_\ad,u^\pm_i)$.  These {\em
analytic superfields} are useful for describing the hypermultiplet as
well as the vector potential superfield for the vector multiplet.
Supersymmetry invariants can be constructed by integrating analytic
superfields against the measure
\be\label{ameas}
\int\! du\, d^4x_A\, d^4\thp = \int\! du\, d^4x\, (D^-)^4 ,
\ee
where $(\Dm)^4$ is defined in (\ref{defDpm4}).

Finally, the reality conditions on superspace actions can be deduced
from the action (\ref{thcc}) and (\ref{ucc}) of complex conjugation on
the coordinates.  In addition, one can introduce another kind of
conjugation, called {\em tilde conjugation}, by combining
complex conjugation with the antipodal map on the 2-sphere.  Thus
\be\label{utc}
\til{\th^i_\a} = -\thb_{\ad i}, \qquad
\til{u^{\pm i}} = - u^\pm_i ,
\ee
from which the tilde conjugation properties of $\th^\pm$, $D^\pm$, and
$D^{\pm\pm}$ can be deduced.  These properties for both complex and
tilde conjugation are summarized in appendix A.4 of \cite{gios01}.
In particular, $\til{\Dp} = -\Dbp$ and $\til{\Dbp} = \Dp$, so that
tilde conjugation preserves analytic superfields.  Notice also that
$\bar{x_C} =\til{x_C}$, while $x_A = \til{x_A}$.

\section{Higgs branch terms}

Hypermultiplets are described by scalar analytic superfields of
$\U(1)$ charge +1, traditionally called $\qp$.  The bosonic terms in
the component expansion of the analytic superfield $\qp$ and its
conjugate $\qtp$ are, in the analytic basis,
\bea\label{qexp}
\qp &=&
F^+ + (\thp)^2 M^- + (\thbp)^2 N^-
+ i \thp \sigma^\mu \thbp A^-_\mu + (\thp)^2 (\thbp)^2 P^{(-3)},
\nonumber\\
\qtp &=&
\til F^+ + (\thp)^2 \til M^- + (\thbp)^2 \til N^-
+ i \thp \sigma^\mu \thbp \til A^-_\mu
+ (\thp)^2 (\thbp)^2 \til P^{(-3)},
\eea
where $F^+$, $M^-$, $N^-$, $A^-_\mu$, and $P^{(-3)}$ are functions
of $x_A^\mu$ and the $u$'s, and the tildes on these functions act as
complex conjugation on the coefficient functions of their
$u$-expansion, while acting as $\SU(2)$ conjugation on the $u$'s as in
(\ref{utc}).  For example, the $u$-expansion of $F^+$ and $\til F^+$
are
\bea\label{Fuexp}
F^+ &=&
f^i(x_A)\, \up_i + f^{(ijk)}(x_A)\, \up_i \up_j \um_k + \cdots ,
\nonumber\\
\til F^+ &=&
\bar f_i(x_A)\, {\up}^i + \bar f_{(ijk)}(x_A)\,
{\up}^i {\up}^j {\um}^k + \cdots
\eea
It turns out, as we will see below, that the $f^i$'s are the propagating
complex scalar fields of the hypermultiplet. As was argued in the
introduction, $f$ (and all propagating scalars, generally) should be
assigned dimension $0$ in the derivative expansion.  Since the
2-sphere coordinates $u^\pm$ were also assigned dimension $0$, we see
that $F^+$ and therefore $\qp$ have dimension $0$.

Since $[d\th]=1/2$, we see that integrating arbitrary functions of the
$\qp$ or $\qtp$ analytic superfields against the analytic measure
(\ref{ameas}) gives a 2-derivative term: $\int\!du\, d^4\thp\,
B(\qp,\qtp)$.  However, since $[u^\pm]=0$, it follows that
$[D^{\pm\pm}] = 0$; also from the algebra of derivatives (\ref{Dalg})
it follows that if $\Phi$ is an analytic superfield, then so is
$(\Dpp)^n \Phi$.  Therefore the general 2-derivative superspace action
is
\be\label{2der}
S^H_2 = \int\! du\, d^4x_A\, d^4\thp \, A(\qp,\qtp; u^\pm,\Dpp) .
\ee
The arbitrary number of $\Dpp$ derivatives that can appear in $S_2$
is an example of the infinite redundancy of the harmonic superspace
formalism, discussed in the introduction.  For the case of the
2-derivative action, this redundancy has been solved in the sense
that it has been shown \cite{gios01} that any 2-derivative action of
hypermultiplets can be realized by $A$'s of the more specific form
\be\label{genhypS2}
A = -\qtp \Dpp \qp + L^{+4}(\qp,\qtp,u^\pm) ,
\ee
where $L^{+4}$ is an arbitrary local functional with no dependence on
the $\Dpp$ derivatives and of total $\U(1)$ charge $+4$.  Note that as
$\qp$ is complex, $\qtp$, which contains the complex conjugate of the
component fields of $\qp$, should be varied independently.
(Alternatively, one could treat $\qp$ and its complex conjugate $\qm$
as independent fields.  Then the form of $S_2$ would be quite
complicated as $\qtp$ is given by a non-local expression involving
an infinite series of $D^{\pm\pm}$ derivatives acting on $\qm$; see,
\eg, eqn.~(3.111) of \cite{gios01}.)  Also, many different
hypermultiplets can easily be included by putting indices on the
$q$'s.  Finally, note that the explicit $u$-dependence in $A$ permits
a non-$SU(2)_R$ invariant action.

Now let us examine the possible superfield form of 3- and 4-derivative
terms in effective actions.  Since the $\thp$ integrations over
the analytic subspace is of dimension 2, to get a 3-derivative term
we must include derivatives acting on the hypermultiplet fields.  The
possibilities are either two spinor covariant derivatives or one
space-time derivative.  However, the supercovariant derivatives either
annihilate the hypermultiplets or don't anticommute with the 
analytic constraints, and so do not give supersymmetry invariants
upon integration over the analytic subspace.  Since space-time 
derivatives commute with all the
supercovariant derivatives, if $\Phi$ is an analytic superfield, then
so is $\del_\mu\Phi$.  But a single space-time derivative cannot
give rise to a Lorentz invariant term.  Thus there are no
3-derivative terms on the Higgs branch.

The observation that $\del_\mu\Phi$ is analytic if $\Phi$ is gives 
a simple way of making
higher-derivative terms from analytic superfields by simply allowing
space-time derivatives in $A$ in (\ref{2der}).  For Lorentz
invariance we need an even number of space-time derivatives, so
the leading term is a 4-derivative term of the form
\be\label{4der-a}
S^H_{4a} =  \int\! du\, d^4x_A\, d^4\thp\,
\del^\mu \qp \del_\mu \qp B(\qp; u^\pm, \Dpp) + \mbox{c.c.},
\ee
for an arbitrary function $B$, where for simplicity we have used $\qp$
to denote either $\qp$ or $\qtp$.  Only $\Dpp$ and not $\Dmm$ can
appear because $\Dmm\qp$ is not analytic.  This type of 4-derivative term
seems to have been missed in other analyses of $N=2$ effective
actions.  A similar term, but with both space-time
derivatives acting on a single field, can always be traded for a term
of the form (\ref{4der-a}) by an integration by parts.

Another analytic combination of superfields and derivatives is
$(\Dp)^4\qm$.  But any action of the form $\int\! du\, d^4x_A
d^4\thp \, (\Dp)^4\qm \, \til B(\qp; u^\pm, \Dpp)$ vanishes
identically since a $\Dp$ derivative annihilates $\til B$, and
so can be taken out of the whole integrand where it is annihilated
by the Grassmann measure.  In fact, any analytic action involving
$(\Dp)^4\qm$'s, and not just the 4-derivative one written
above, vanishes for the same reason.

Other higher-derivative terms can arise from Lagrangians which
do not obey the analytic constraint.  To be supersymmetry invariants,
these actions must then be integrated over all of harmonic superspace.
The integrand can then contain an arbitrary function of any of the
derivatives, as well as $u^\pm$, $\qp$ and its complex conjugate
(anti-analytic) superfield $\qm \equiv \bar{\qp}$ (as well as their
tilde-conjugates).  Since the measure $du\,d^4\thp\,d^4\thm$ already
has dimension four, the most general non-analytic 4-derivative term
has the form
\be\label{4der-b}
S^H_{4b} =  \int\! du\,d^4x_A\, d^8\th\,
\Gamma(\qp,\qm; u^\pm, D^{\pm\pm}) ,
\ee
for an arbitrary function $\Gamma$, where again we use $q^\pm$
to denote also ${\til q}^\pm$.  

An important point in the harmonic superspace formalism reviewed above
is that the auxiliary 2-sphere coordinates $u^\pm$ are not physical
coordinates: they are always integrated over in any physical quantity.
Therefore in writing harmonic superspace Lagrangians there is no
constraint of locality with respect to the $u^\pm$ coordinates.  
For example, terms containing both fields and
tilde-conjugated fields---which involve a non-local inversion on the
2-sphere---are allowed; also 2-sphere non-locality can appear through
terms having arbitrarily large numbers of $D^{\pm\pm}$
derivatives, or multiple $du$ integrations.  Indeed, for a given
dimension term in an effective action expansion, like (\ref{4der-a}) and
(\ref{4der-b}) above, there are infinitely many allowed harmonic superspace 
terms since there is no restriction on the number of $D^{\pm\pm}$'s. This
non-locality is a major technical obstacle to using the harmonic
superspace formalism for making systematic derivative expansions of
effective actions.  
Nevertheless, after integrating over the 2-sphere and removing all the
auxiliary fields with their equations of motion, there remain only a
finite number of distinct terms of a given dimension.  Thus the
2-sphere non-locality in harmonic superspace represents a redundancy
in its description of supersymmetric Lagrangians.  In practice, this
infinite redundancy means that there are infinitely many different
dependencies of the action on the infinitely many auxiliary fields of
harmonic superfields.  But all of these actions reduce to the same
action when the auxiliary fields are substituted in terms of
propagating fields using their equations of motion.

Now, as was discussed in the introduction, once the form of the
leading term in the derivative expansion of the effective action is
fixed, then there is a systematic procedure to compute the corrected
equations for the auxiliary fields order-by-order, implying that the
redundancy in the form of harmonic superspace actions can be
circumvented in principle, thus allowing a systematic classification and
construction of higher-derivative terms in effective actions.  The key
point for our purposes is that given the 2-derivative term,
$S_2$, of the superspace effective action, the 4-derivative terms
are given by a dimension four superspace effective action $S_4$
{\em with auxiliary field components evaluated at their values given
by $S_2$}, as in (\ref{onshellact}).  

In the rest of this section we use this understanding 
to prove that no WZ terms can be constructed on the Higgs branch
and also to show that the appearance of arbitrary combinations
of $D^{\pm\pm}$'s in (\ref{4der-a}) and (\ref{4der-b}) can be 
brought under control.

A WZ term is a 4-derivative term where the propagating scalars
$\phi^a$ enter in the Lagrangian as
\be\label{WZform}
\lambda_{abcd}(\phi) \ep^{\mu\nu\rho\si} \del_\mu \phi^a
\del_\nu \phi^b \del_\rho \phi^c \del_\si \phi^d ,
\ee
with some (generally singular) antisymmetric coefficient
function $\lambda_{abcd}$.  So we wish to search for 
4-derivative terms $S_4$ made from hypermultiplet superfields
in harmonic superspace that can give rise to (\ref{WZform})
in their component expansion after substituting out all the
auxiliary fields.
It is immediately clear that terms of the form (\ref{4der-a}) cannot give rise 
to WZ terms:  when expanded in terms of propagating scalars, two
of the four space-time derivatives must come from the explicit
space-time derivatives in (\ref{4der-a}); however since they are
contracted, they cannot contribute to a WZ term (\ref{WZform})
where the derivatives are all antisymmetrized.  Thus we search
for WZ terms among terms of the form (\ref{4der-b}).

\subsection{Free hypermultiplets}

We start with free hypermultiplets to illustrate our argument in
an algebraically simple setting, and later we will generalize to
arbitrary hypermultiplet 2-derivative actions.

The free hypermultiplet action is \cite{gios01}
\be\label{freehyp}
S^H_2 = -\int\!\! du\,d^4x_A\, d^4\thp\, \qtp \Dpp \qp .
\ee
Using the fact that in the analytic basis $\Dpp =\del^{++}  
-2i \thp \sigma^\mu \thbp \del_\mu+\cdots$ (where we've defined 
$\del^{++}\equiv u^{+i}\del/\del u^{-i}$), that $\int
d^4\thp\, (\thp)^2 (\thbp)^2 = 1$, and using the identity $(\thp \sigma^\mu
\thbp) (\thp \sigma^\nu \thbp) ={1\over2}\eta^{\mu\nu} (\thp)^2
(\thbp)^2$, we find the bosonic components of (\ref{freehyp}) are
\bea
S^H_2 &\supset&
-\!\int\!\! du\,d^4x \Bigl[
\til F^+ \left( \del^{++} P^{(-3)} + \del^\mu A^-_\mu \right)
+ \til M^- \del^{++} N^-
+ \til N^- \del^{++} M^-
\nonumber\\
&& \qquad \qquad \quad
{}+ \til A^{\mu-} \left( \del_\mu F^+ - {\textstyle{1\over 2}}
\del^{++} A^-_\mu \right) + \til P^{(-3)} \del^{++} F^+
\Bigr] .
\eea
Varying with respect to the tilded fields one finds algebraic equations of
motion whose solutions are simply
\be\label{eomsolns}
F^+ = f^i(x_A)\, \up_i,\qquad
A^-_\mu = 2 \del_\mu f^i(x_A)\, \um_i, \qquad
M^-=N^-=P^{(-3)}=0 ,
\ee
plus the free equation of motion $\del^2 f^i=0$ (coming from the lowest
$u$-component of the $\til F^+$ equation).  Thus all the fields except
the first component of the $u$-expansion of $F^+$ are auxiliary fields.

Since we are only interested in extracting the purely bosonic
4-derivative terms from $S^H_{4b}$, and given the result from the
introduction that we need substitute the auxiliary fields in $S^H_{4b}$
using their 2-derivative values, it follows that {\em it is sufficient
to use the $\qp$ superfield modulo the constraints
(\ref{eomsolns})}.  It will be convenient to deal with the various
complex and tilde conjugates of $\qp$ in parallel.  Denote by $\qm$
the complex conjugate of $\qp$, so the various conjugates satisfy
\be\label{constraints}
0 = \Dp \qp = \Dbp \qp ,\quad
0 = \Dp \qtp = \Dbp \qtp ,\quad
0 = \Dm \qm = \Dbm \qm ,\quad
0 = \Dm \qtm = \Dbm \qtm .
\ee
Their bosonic component expansions modulo the free action
auxiliary field equations of motion (\ref{eomsolns}) are then,
in the analytic basis,
\bea\label{qanalexp}
\qp &=& +f^i(x_A)\, \up_i + 2i (\thp \!\!\!\not\!\del \thbp)
f^i(x_A)\, \um_i ,
\nonumber\\
\qtp &=& +\fb_i(x_A)\, {\up}^i + 2i (\thp \!\!\!\not\!\del \thbp)
\fb_i(x_A)\, {\um}^i ,
\nonumber\\
\qm &=& -\fb_i(\bar x_A)\, {\um}^i + 2i (\thm \!\!\!\not\!\del \thbm)
\fb_i(\bar x_A)\, {\up}^i ,
\nonumber\\
\qtm &=& +f^i(\bar x_A)\, \um_i - 2i (\thm \!\!\!\not\!\del \thbm)
f^i(\bar x_A)\, \up_i .
\eea
It will be convenient to expand the expressions (\ref{qanalexp}) in the
central basis where the full $\th$ dependence is manifest.
Using (\ref{xa}) and defining the shorthands
\bea\label{fpmdefs}
\dpp \equiv \thp \!\!\!\not\!\del \thbp ,\qquad
\dpm \equiv \thp \!\!\!\not\!\del \thbm ,&&
\dmp \equiv \thm \!\!\!\not\!\del \thbp ,\qquad
\dmm \equiv \thm \!\!\!\not\!\del \thbm ,\nonumber\\
f^\pm \equiv f^i(x)u^\pm_i, &&
\fb^\pm \equiv \fb_i(x) u^{\pm i},
\eea
we find
\bea\label{qcenexp}
\qp &=& +\fp + 2i \dpp \fm - i \dpm \fp - i \dmp \fp
+ {\cal O}(\del^2f) ,
\nonumber\\
\qtp &=& +\fbp + 2i \dpp \fbm - i \dpm \fbp - i \dmp \fbp
+ {\cal O}(\del^2f) ,
\nonumber\\
\qm &=& -\fbm + 2i \dmm \fbp - i \dpm \fbm - i \dmp \fbm
+ {\cal O}(\del^2f) ,
\nonumber\\
\qtm &=& +\fm - 2i \dmm \fp + i \dpm \fm + i \dmp \fm
+ {\cal O}(\del^2f) ,
\eea
where ${\cal O}(\del^2f)$ stands for terms with two or more space-time
derivatives acting on $f$.  We can neglect the 2-derivative terms
acting on a single field since those derivatives are necessarily
symmetrized and so can never contribute to the WZ term (\ref{WZform}).

Since in the central basis $D^{\pm\pm}$ act simply as $D^{\pm\pm} u^\pm
=0$ and $D^{\pm\pm}u^\mp =u^\pm$, we get from (\ref{qcenexp})
\bea\label{qexpid}
\Dpp\qp &=& {\cal O}(\del^2f) ,
\nonumber\\
\Dpp\qm &=& -\fbp - 2i \dpp \fbm + i \dpm \fbp + i \dmp \fbp
+ {\cal O}(\del^2f) \, = \, -\qtp + {\cal O}(\del^2f) ,
\nonumber\\
\Dmm\qp &=& +\fm - 2i \dmm \fp + i \dpm \fm + i \dmp \fm
+ {\cal O}(\del^2f) \, = \, +\qtm + {\cal O}(\del^2f) ,
\nonumber\\
\Dmm\qm &=& {\cal O}(\del^2f) ,
\eea
where we used $[\Dpp,\dpm]= \dpp$ \etc.  This means that as far as the WZ
terms are concerned, we need only consider the four fields $q^\pm$ and
$\qt^\pm$, since $D^{\pm\pm}$ acting on them gives back the same four fields
up to higher space-time derivative terms when the auxiliary fields are 
put on shell.  This is an example of the identities relating
superfields evaluated at $a=a_2$ to some number of their $\del/\del u$
derivatives referred to in the introduction: (\ref{qexpid}) shows that
all derivatives of hypermultiplet fields can be reduced to one of four
possibilities as far as the WZ term is concerned.  

Thus, with complete generality, we can take any potential
hypermultiplet WZ term to be of the form
\be\label{4der}
S_4 =  \int\! du\,d^4x\,d^8\theta \, \Gamma(\qp,\qm,\qtp,\qtm; u^\pm) ,
\ee
for an arbitrary real function $\Gamma$, with no $D^{\pm\pm}$
dependence.  Note that $\qt^\pm$ should not be thought of as fields
to be varied independently of $q^\pm$.

More generally, if we wanted to
classify all the different non-WZ 4-derivative terms, we would
have to include more---though still a finite number---possibly
distinct combinations of $u$-derivatives on fields.  Indeed, 
it is not hard to show\footnote{Including the fermions and using the
2-derivative equations of motion for the auxiliary fields, we have, in the 
analytic basis, $\qp = \fp + 2i\dpp\fm + \thp\psi + \thbp\psib$.  In the 
analytic basis $\Dpp = \del^{++} + \th^{+\a}\del^+_\a +
\thb^{+\ad}{\bar\del}^+_\ad - 2i\dpp$.  Direct computation then gives
$\Dpp\qp=4(\dpp)^2\fm-2i\dpp(\thp\psi+\thbp\psib)$, $(\Dpp)^2\qp
= 4(\dpp)^2\fp$, and thus $(\Dpp)^3\qp=0$.} that, using the auxiliary 
component equations of motion, $(\Dpp)^3\qp=0$.  This implies that
in a central basis expansion of (\ref{qanalexp}), keeping
all the higher-derivative as well as fermionic terms, the 
$u$-expansion of $\qp$ contains only the three terms $\up$, $(\up)^2 \um$,
and $(\up)^3 (\um)^2$.  So $(\Dmm)^4\qp=0$ as well.  Thus there are only
six non-vanishing combinations of $D^{\pm\pm}$ derivatives 
acting on $\qp$.  For the holomorphic
term, (\ref{4der-a}), only $\qp$ and $\Dpp$ appear, so there are only
three combinations to consider,
\be\label{3comb}
\qp, \Dpp\qp, (\Dpp)^2\qp ,
\ee
while for the non-holomorphic term, (\ref{4der-b}), where $q^\pm$ and
$D^{\pm\pm}$ may appear there are twelve non-vanishing combinations
\bea\label{12comb}
&&(\Dmm)^3\qp,\ (\Dmm)^2\qp,\ \Dmm\qp,\ \qp,\ \Dpp\qp,\ (\Dpp)^2\qp ,
\nonumber\\
&&(\Dpp)^3\qm,\ (\Dpp)^2\qm,\ \Dpp\qm,\ \qm,\ \Dmm\qm,\ (\Dmm)^2\qm .
\eea
Only the three combinations (\ref{3comb}) need be considered in 
(\ref{4der-a}), and only the twelve combinations (\ref{12comb}) 
need be considered in  
(\ref{4der-b}).  This leads to a finite classification of 4-derivative
terms on the Higgs branches.

Returning to our search for the WZ term, we expand (\ref{4der}) by 
inserting the expressions (\ref{qcenexp}) and keeping four of the derivative
terms.  To survive the $d^8\th$ integration and to get a
tr$(\sigma^\mu \bar\sigma^\nu \sigma^\rho \bar\sigma^\sigma)$ (so that
we get an $\ep^{\mu\nu\rho\sigma}$) we need one of each type of
derivative term: $\dpp\dpm\dmm\dmp$.  From (\ref{qcenexp}), we
see that for every field the $\dpm$ and $\dmp$ contributions always
enter together in the form
\be
(\dpm+\dmp)A ,
\ee
where $A$ stands for some field.  Therefore in the expansion of
(\ref{4der}), contributions to potential WZ terms will always appear in
the combination
\be
\int\!du\, d^4x\, d^8\th\, (\dpm+\dmp)A \ 
(\dpm+\dmp)B \  \dpp C \  \dmm D ,
\ee
for some $A$, $B$, $C$, and $D$.  Doing the $\th$ integrals and
keeping only the $\ep^{\mu\nu\rho\sigma}$ piece from the $\sigma$
trace, we have
\be
\int\!du\, d^4x\, \ep^{\mu\nu\rho\sigma}
(\del_\mu A \del_\nu B + \del_\nu A \del_\mu B)
\del_\rho C \del_\sigma D ,
\ee
which vanishes by antisymmetry.  Therefore all the potential WZ terms
vanish for free hypermultiplets.

Actually, we did not need the identities (\ref{qexpid}) to reach this
conclusion.  It is enough to note that in $\qp$ the derivatives of
the scalar fields enter only in the three combinations $\dpp f$,
$\dmm f$, and $(\dpm+\dmp) f$, which form a triplet $\SU(2)$ representation.
Thus these combinations close among themselves under $u$-differentiation,
\bea
D^{\pm\pm} (\dpm+\dmp) &=& 2 \not\!\del^{\pm\pm},\nonumber\\
D^{\pm\pm} (\not\!\del^{\pm\pm}) &=& 0,\nonumber\\
D^{\pm\pm} (\not\!\del^{\mp\mp}) &=& \dpm+\dmp. \eea It follows
that only these combinations can occur in the expansion of the
most general 4-derivative action (\ref{4der-b}), and so, by 
arguments of the previous paragraph, cannot give rise to a WZ term.

Alternatively, one can check this argument by a direct
calculation to extract a WZ-type term from (\ref{4der}). 
Up to a total space-time derivative, the $\th$ integrations
can be replaced by supercovariant differentiation evaluated
at $\th=0$.  This differentiation gives a total derivative
for the WZ-like terms:
\bea 
S_4 &=&{1\over16}\int\! du \,d^4x\, 
\left[ \Gamma_{\Ibar\Jbar KL}
\,{\Dbp}^\ad {\Dp}^\a \qm_I\ \Dbp_\ad {\Dm}^\b \qp_L\ 
{\Dbm}^\bd\Dm_\b \qp_K\ \Dbm_\bd \Dp_\a \qm_J 
\right]_{\th=0}
\nonumber\\ 
&=& \int\! du \,d^4x\,
\Gamma_{\Ibar\Jbar KL}
\,\ep^{\mu\nu\rho\si}\,\del_\mu \fp_L\,\del_\nu \fm_K\,
\del_\rho\bar\fm_\Ibar\,\del_\si\bar\fp_\Jbar
\nonumber\\
&=& \int\! du \,d^4x\,\,\del_\mu
\left(\Gamma_{\Ibar\Jbar K}
\ep^{\mu\nu\rho\si}\, \del_\nu \fm_K\, \del_\rho 
\bar\fm_\Ibar\, \del_\si \bar\fp_\Jbar\right) ,
\eea
where the subscripts denote differentiation of $\Gamma$ with
respect to its arguments.  We are using a notation, to be introduced
shortly, in which subscripts $I,J,K,L$ label both the hypermultiplets
and their tilde conjugates, so that the indices run from 1 to $2n$ where
$n$ is the number of hypermultiplets.

\subsection{General hyperkahler geometry}

We now extend the above argument to general 2-derivative terms
for hypermultiplets (\ref{genhypS2}) following the notation of
section 11.4 of \cite{gios01}.  Consider a theory with $n$
massless neutral hypermultiplets $q^+_I$, $I=1,\ldots,n$.  Instead
of treating the $\til q^+_I$'s separately, it is convenient to
double the number of fields, letting $I$ run from 1 to $2n$, and
to impose the condition
\be\label{sp2ncond}
\til{q^+_I} \equiv q^{+I} = \Omega^{IJ} q^+_J ,
\ee
where $\Omega^{IJ}$ is the antisymmetric $\Sp(2n)$ invariant
tensor which has the matrix form ${0\ -1\choose 1\ \ 0}$ in
$n\times n$ blocks.  Then the general hypermultiplet 2-derivative
action can be written
\be\label{ghypS2}
S^H_2 = \int\! du\,d^4x_A\,d^4\thp\, \textstyle{1\over2}
\left( q^+_I \Dpp q^{+I} + L^{(+4)}(q^+,u^\pm) \right),
\ee
where $L^{(+4)}$ is an arbitrary function of the $q^+$'s and the
$u^\pm$'s.

Inserting the $\th$-expansion (\ref{qexp}) of $q^+$ and doing
the $\th$ integration one finds the bosonic part of the action to be
\bea
S^H_2 &=& \int\! du\,d^4x_A \Bigl[
F_I (\del^{++} P^I + \del^\mu A^I_\mu)
- \textstyle{1\over4} A^\mu_I \del^{++} A_\mu^I
+ M_I \del^{++} N^I \nonumber\\
&& \qquad\qquad\qquad\quad
{}+ \textstyle{1\over2} P^I \del_I L
- \textstyle{1\over8} (A^{\mu I}A^J_\mu -4 M^I N^J)
\del_I\del_J L \Bigr] ,
\eea
where we have dropped the $\U(1)$ charge superscripts on the $F^+$,
$A^-_\mu$, $M^-$, $N^-$, and $P^{(-3)}$ component fields to reduce
clutter, and where $L = L^{(+4)}(F^+,u^\pm)$ and $\del_a =
\del/\del F^{+a} $.  The equations of motion following from this
action are
\bea\label{gheom}
\del^{++} F_I &=& \textstyle{1\over2} \del_I L , \nonumber\\
\mbox{}{{\cal D}^{++}}^I_J M^J &=& {{\cal D}^{++}}^I_J N^J = 0,\nonumber\\
\mbox{}{{\cal D}^{++}}^I_J A_\mu^J &=& 2 \del_\mu F^I,\nonumber\\
\mbox{}{{\cal D}^{++}}^I_J P^J &=& - \del^\mu A_\mu^I - \textstyle{1\over8}
(\del^I\del_J\del_K L) (A^{\mu J}A_\mu^K - 4 M^J N^K) ,
\eea
where we have 
defined ${{\cal D}^{++}}^I_J =  \de^I_J \del^{++} -
{1\over2} \del^I \del_J L$.
As in the free case, the leading term in the $u$-expansion of $F^I$,
$f^{Ii}(x_A) u^+_i$, contains the propagating fields, while all the
other components are auxiliary.  They are determined by the above
equations in terms of the $f^{Ii}$:
\bea\label{ghsolns}
F^I &=& f^{Ii} u^+_i + V^I(f,u),\nonumber\\
M^I &=& N^I = 0,\nonumber\\
A^I_\mu &=& -2 E^I_{Ji} \del_\mu f^{Ji}, \nonumber\\
P^I &=& G^I_{Ji,Kj} \del_\mu f^{Ji} \del^\mu f^{Kj}
+ H^I_{Ji} \del^\mu\del_\mu f^{Ji},
\eea
where the $V^I(f,u)$ $u$-expansion starts with $\up_{(i}\up_j\um_{k)}$
and is determined by the first equation in (\ref{gheom}).  Here
$E^I_{Ji}(f,u)$ is determined by the equation ${{\cal D}^{++}}^I_J 
E^J_{Ki} = -{\del F^I/\del f^{Ki}}$,
and $G(f,u)$ and $H(f,u)$ are determined by similar
differential equations in $u$.  Though explicit expressions
for $F$, $E$, $G$, and $H$ might be difficult to find for
a given $L$, they are local functionals of the
scalar fields $f^{Ii}$, but not of their derivatives.

By our general arguments from section 1, to find the purely
bosonic 4-derivative terms coming from $S^H_{4b}$ (\ref{4der-b}),
it is sufficient to substitute $q^+$ modulo the constraints
(\ref{ghsolns}).  Furthermore, we can neglect $P^I$ since
it is proportional to 2-derivative terms with Lorentz
indices contracted.  This can never contribute to the
WZ term (\ref{WZform}).  Thus, the bosonic components
of $q^+_I$ and their complex conjugates $q^-_I$ are
(we are putting the $U(1)$ charge superscripts back on
$F^+$ and $E^-$ now)
\bea
q^{+I} &=& F^{+I}(f,u) -2i {E^-}^I_{Ji}(f,u) \dpp f^{iI}
+ {\cal O}(\del^2f) , \nonumber\\
q^{-I} &=& F^{-I}(\bar f,u) +2i {E^+}^{Ii}_J(\bar f,u) \dmm \bar f^I_i
+ {\cal O}(\del^2f) ,
\eea
where we have defined $F^- = \bar{F^+}$ and $E^+ = \bar{E^-}$.
So far we have been working in the analytic basis, where $f^{iI}$ is
a function of $x^\mu_A$.  Using (\ref{xa}), we convert to the
central basis where the full $\th$ dependence is manifest:
\bea
q^{+I} &=& F^{+I} -2i {E^-}^I_{Ji} \dpp f^{iJ}
+ i {{\cal D}^{++}}^I_J {E^-}^J_{Ki} (\dpm+\dmp) f^{iK}
+ {\cal O}(\del^2f) , \nonumber\\
q^{-I} &=& F^{-I} +2i {E^+}^{Ii}_J \dmm \bar f^J_i
-i {{\cal D}^{--}}^I_J {E^+}^{Ji}_K (\dpm+\dmp) \bar f^K_i
+ {\cal O}(\del^2f) ,
\eea
where we have used the definition of ${E^-}^I_{Ji}$ and 
its complex conjugate.

Notice that $q^\pm$ depend on the derivatives of the
scalars only through the $\SU(2)$ triplet combinations
$\not\!\del^{\pm\pm}$ and $\dpm+\dmp$ just as in the
free hypermultiplet case.  Then the argument of the previous 
subsection again shows that no WZ term can be generated 
with hypermultiplets alone.

\section{Coulomb branch terms}

The unconstrained $N=2$ vector multiplet superfield is a
$\U(1)$-charge +2 analytic superfield $\Vpp$, satisfying a reality 
condition
\be
\til\Vpp = \Vpp
\ee
and transforming
under $\U(1)$ gauge transformations as 
\be\label{devpp}
\de\Vpp = - \Dpp \lambda,
\ee
where $\lambda$ is an arbitrary real ($\til\lambda=\lambda$) 
analytic superfield of $\U(1)$-charge 0.  Though both $\lambda$ and 
$\Vpp$ have infinite $u$ expansions, we can use the gauge
freedom (\ref{devpp}) to eliminate all but a finite number of the 
components of $\Vpp$ (an analog of the Wess-Zumino gauge in 
$N=1$ supersymmetry):
\bea\label{Vppexp}
\Vpp &=& 
i\sqrt2 \phi(x_A) (\thbp)^2 
- i\sqrt2 \bar\phi(x_A) (\thp)^2
- 2i A_\mu(x_A) \thp\sigma^\mu\thbp
\\
&&\ \mbox{}
+ 4(\thbp)^2 \th^{+\a} \psi^i_\a(x_A) u^-_i
- 4(\thp)^2 \thbp_\ad \bar \psi^{\ad i}(x_A)u^-_i 
+ 3(\thp)^2 (\thbp)^2 D^{ij}(x_A)u^-_i u^-_j,
\nonumber
\eea
where $D^{ij}$ are real scalars, $\phi$, $\psi^i_\a$, and $D^{ij}$
are gauge invariant, and the real vector $A_\mu$ transforms under a
residual gauge invariance in the usual way as $\de A_\mu = \del_\mu 
\ell$ for $\ell$ an arbitrary real function.

\subsection{The field strength superfield in $N=2$ superspace}

The gauge invariant field strength superfield is constructed as follows.
First, another gauge potential superfield $\Vmm$ is defined in terms 
of $\Vpp$ as the solution to the differential equation in $u^\pm$ 
\be\label{Vmmdef}
\Dpp \Vmm=\Dmm \Vpp ,
\ee
which has a unique solution by virtue of the harmonicity requirement on
the $u$-sphere.  $\Vmm$ is not an analytic (or anti-analytic) 
superfield, but is real $\Vmm=\til\Vmm$ and transforms under gauge 
transformations as $\de\Vmm = -\Dmm\lambda$.
The field strength superfield is then defined by
\be\label{Wdef}
W=-{1\over 4 }(\Dbp)^2 \Vmm .
\ee
It is a straight forward exercise, using the $N=2$ algebra (\ref{su2alg}) and
(\ref{Dalg}), to check that $W$ is gauge invariant, chiral
\be
\Db^\pm W =0,
\ee
satisfies the Bianchi identities
\be\label{bianchi}
D^{i\a} D^j_\a W = \Db^i_{\ad} \Db^{j\ad} \Wb ,
\ee
and is $u$-independent
\be
D^{\pm\pm} W = 0 .
\ee
Thus, in expressions involving the field strength superfields alone
(\ie, no $V^{\pm\pm}$'s), the integration over the auxiliary $u$-sphere
can be done separately, leaving an expression in standard $N=2$ superspace
with coordinates $\{x^\mu, \th^\pm_\a, \thb^\pm_\ad\}$.

Thus $W=W(x_C,\th^\pm)$, and the component expansion of $W$ starts with 
the complex scalar $\phi$ introduced in (\ref{Vppexp}):
\be
W(x_C,\th^\pm) = i\sqrt2 \phi(x_C)+\cdots .
\ee
Thus the derivative dimensions of $W$ and $V^{\pm\pm}$ are
\be
[W]=0, \qquad [V^{\pm\pm}]=-1,
\ee
where the second follows from (\ref{Vmmdef}) and (\ref{Wdef}).

If we assume that all $N=2$ supersymmetric expressions on the Coulomb
branch can be written solely in terms of the field strength superfield
$W$, its complex conjugate, and derivatives, and is local in $N=2$
superspace, then the general form of higher-derivative terms in
the effective action is easy to obtain.  So, in the remainder of this subsection
we will make these two assumptions, and classify the terms up to
four derivatives.  But, in the next subsection we will revisit these
assumptions and find that the interplay between gauge invariance and
locality in superspace is algebraically complicated.

With these simplifying assumptions, the leading term in
the derivative expansion of the low energy effective action
is the 1-derivative Fayet-Iliopoulos term
\be
S^C_1 = \int\! d^4x\, d\th^i\cdot d\th^j \xi_{ij} W + \mbox{c.c.} ,
\ee
where $\xi_{ij}$ are an $\SU(2)$ triplet of real constants.
Though this is an integral over only 1/4 of superspace, it
is $N=2$ invariant by virtue of the extra constraint (\ref{bianchi})
that $W$ satisfies.  This constraint does not lead to any
other local supersymmetry invariants, so higher-derivative
terms can be constructed by treating $W$ as an unconstrained
chiral superfield.  Then the general 2-derivative term is
the well-known holomorphic pre-potential term given by an integral
over the chiral half of superspace,
\be
S^C_2 = \int\! d^4x\, d^4\th\ {\cal F}(W) +\mbox{c.c.} .
\ee

In close analogy to our discussion of the 3- and 4-derivative
terms for the hypermultiplets in the paragraphs following 
(\ref{genhypS2}), but for chiral fields instead of analytic
ones, we find that there are no 3-derivative terms, and two
independent 4-derivative terms:
\bea\label{CBterms}
S^C_{4a} &=& \int\! d^4x\, d^4\th \,\,
\del_\mu W \del^\mu W \, {\cal G}(W) +\mbox{c.c.},
\nonumber\\
S^C_{4b} &=& \int\! d^4x\, d^4\th\, d^4\thb \,\,
{\cal H}(W,\Wb) .
\eea
Unlike the hypermultiplet case, since there is no $u$-dependence
in these terms, there is no redundancy coming from arbitrary
$D^{\pm\pm}$ derivatives.  The holomorphic $S^C_{4a}$ term
seems to have been ignored in the literature.  

Note that the subset of $S^C_{4a}$-type terms which can be written using
integration by parts as $\int\!d^4x d^4\th\, \del^2W\, {\cal J}(W)$
are actually special cases of $S^C_{4b}$ terms, by virtue of the 
constraints (\ref{bianchi}). 
This follows because for an $S^C_{4b}$ term with $\cal H$ of the 
special form $\Wb {\cal J}(W)$ we have
\bea
4 \int\!d^4x\, d^8\th\ \Wb\, {\cal J}(W) 
&=& {1\over4} \int\!d^4x\, d^4\th\ 
\left[(\Dbp)^2 (\Dbm)^2 \left( \Wb\, {\cal J}(W)
\right) \right]_{\thb=0}
\\
&=& {1\over4} \int\!d^4x\, d^4\th\ {\cal J}(W)\, 
\left[(\Dbp)^2 (\Dbm)^2 \Wb\right]_{\thb=0} 
\nonumber\\
&=& {1\over4} \int\!d^4x\, d^4\th\ {\cal J}(W)\, 
\left[(\Dbp)^2 (\Dm)^2 W \right]_{\thb=0} 
\nonumber\\
&=& 2\int\!d^4x\, d^4\th\ {\cal J}(W)\, 
\left[\del^2 W \right]_{\thb=0} 
= \int\!d^4x\, d^4\th\ {\cal J}(W)\, \del^2 W + \mbox{c.c.}
\nonumber
\eea
In the first line we replaced the antichiral integrations by supercovariant 
derivatives evaluated at $\thb=0$; in the second line we used the chirality 
of $W$; in the third line we used the Bianchi identity (\ref{bianchi}) in
the form $(\Dbm)^2\Wb=(\Dm)^2 W$; and in the last line we used the
supersymmetry algebra (\ref{Dalg}) to commute the $\Dbp$'s past the
$\Dm$'s.  Examples of $S^C_{4a}$ terms which cannot be rewritten in this
way as $S^C_{4b}$ terms require two or more vector multiplets.

In the search for $N=2$ supersymmetric WZ terms, it is clear
that they will not be found in $S^C_{4a}$ since two of
the space-time derivatives are contracted, ruling out terms such
as (\ref{WZform}) antisymmetrized on derivatives of scalars.
The remaining possibility is the integral expression over the whole
superspace of the form $S^C_{4b}$.
An expansion of $W$ in the central basis up to first derivatives of the
scalar field $\phi$ gives
\be\label{Wphiexp}
W= i\sqrt2 \phi + \sqrt2 (\dpm-\dmp)\phi+{\cal O}(\partial^2).
\ee 
Indeed, the form of this expansion can be deduced without any calculation,
since only $\dpm$ and $\dmp$, and not $\dpp$ and $\dmm$, can appear,
because $W$ has a vanishing $\U(1)$ charge, while the $u$-independence of 
$W$ implies that only the antisymmetric combination $(\dpm-\dmp)$ of derivatives
can occur.  As we mentioned in the hypermultiplet case, in
order to get the $\ep^{\mu\nu\rho\sigma}$ tensor required
for a WZ term, we need one of each type of derivative term $\dpp\phi$,
$\dpm\phi$, $\dmm\phi$, and $\dmp\phi$.  Since only one independent
combination of those four derivatives appears in (\ref{Wphiexp}),
we conclude that $S^C_{4b}$ cannot contain a WZ term and thus that
there is no WZ term on the Coulomb branch.

\subsection{Superspace Chern-Simons-like terms and Grassmann non-locality}

The above conclusions only hold modulo the two assumptions we made:
(a) manifest gauge invariance and (b) Grassmann locality.
Manifest gauge invariance means that all Coulomb branch terms can
be written solely in terms of the
field strength superfield $W$.  Grassmann locality means that these
terms are local in the $N=2$ superspace Grassmann-odd coordinates.
In this subsection we will examine these assumptions, and will give
simple arguments showing that either manifest gauge invariance
or Grassmann locality holds, but that to show both simultaneously
involves a case by case analysis at each order of the derivative expansion.
It is interesting to note that this problem has nothing to do with
harmonic superspace, and exists as well for $N=1$ supersymmetric theories.

First, let's consider the issue of (a) manifest gauge invariance.
The question is whether there exist \emph{superspace Chern-Simons-like} 
terms,
that is, terms in the effective action which are gauge invariant, but
that cannot be written solely in terms of the gauge invariant field
strength superfield $W$ and its derivatives, and must also
involve the potential superfield $V$.  (The following arguments work
as well for $N=1$ as $N=2$ supersymmetry, so we drop the indices
on $V$ and $W$; for $N=1$ supersymmetry, $V$ is a real scalar superfield,
and $W_\a$ is a chiral spinor superfield, while for $N=2$ we have seen
that $\Vpp$ is a real analytic scalar superfield, while $W$ is a chiral
scalar superfield.)  Consider the general expression for a term in the
effective action involving vector multiplets, schematically:
\be\label{genvecterm}
S^C = \int\! d\zeta\, f(V, D) ,
\ee
where $d\zeta$ is the measure on the appropriate superspace and $D$
denotes all the various covariant derivatives.  A partial fixing of
the gauge invariance (for either $N=1$ or $N=2$ vector multiplets) allows
us to set all but a finite number of auxiliary fields to zero, leaving
the gauge-variant vector potential, $A_\mu$, as well as gauge
invariant scalars and spinors, which we'll collectively denote by $\phi$, 
as component fields.  In this gauge we have
\be\label{genvecterm2}
S^C = \int\! d^4x\, g(A_\mu, \phi, \del_\nu),
\ee
where $g$ is Lorentz invariant and gauge invariant under $\de A_\mu = 
\del_\mu \ell$.  Since the $\phi$'s are gauge invariant and there are
no Chern-Simons-like terms (as opposed to \emph{superspace} Chern-Simons-like) 
terms in even dimensions,\footnote{Actually, we
do not know of a proof of this ``folk theorem'' which states that
for every gauge-invariant $f$ there exists a $g$ such that $\int\!d^{2n}x\, 
f(A_\mu,\del_\nu) =  \int\!d^{2n}x\, g(F_{\mu\nu},\del_\rho)$ (modulo
surface terms).} it follows 
that up to total derivatives (\ref{genvecterm2}) can be written as
\be\label{genvecterm3}
S^C = \int\! d^4x\, h(F_{\mu\nu}, \phi, \del_\rho) .
\ee
Finally, $F_{\mu\nu}$, $\phi$, and their derivatives are
just components of the field strength superfield $W$ and its derivatives,
so we can write
\be\label{genvecterm4}
S^C = \int\! d^4x\, h\left(\textstyle{\int\!d\th_1 j_1(W,D)\,,
\,\int\!d\th_2 j_2(W,D)\,,\ldots}\right) ,
\ee
where the $j_n$ are arbitrary functions of superspace covariant
derivatives and $W$'s, and the $d\th_i$ are appropriate integration
measures over the Grassmann-odd superspace coordinates.  Thus we have 
rewritten the general vector multiplet term (\ref{genvecterm}) solely in 
terms of the field strength superfield.  

But (\ref{genvecterm4}) is not local in superspace.  Such a superspace-local
term would have just a single integral over the Grassmann-odd coordinates,
\be\label{locvec}
S^C_{\mbox{local}} = \int\! d^4x\,d\th \,\, h(W,D) .
\ee
This brings us to the issue of (b) Grassmann locality.  Since the Grassmann-odd
$\th$'s are not physical coordinates, there is no {\em a priori} reason
that effective actions should be local in the $\th$'s.  However,
for unconstrained superfields, locality in space-time
together with supersymmetry invariance actually imply locality in the $\th$'s.
This argument, which is a basic reason for the usefulness of superspace, is 
reviewed in appendix A.  Nevertheless, this observation does not allow us
to conclude that the general vector multiplet term (\ref{genvecterm4})
can be written in the local form (\ref{locvec}), since the field strength
superfield $W$ is constrained by the Bianchi identities
and thus the locality argument does not work.

Thus, our arguments leave open the possibility that there may exist
supersymmetric terms in the effective action for vector multiplets (for
$N=1$ as well as $N=2$ supersymmetry) which can only be written in the 
Grassmann non-local form (\ref{genvecterm4}).  We call such terms
superspace Chern-Simons-like terms since, like Chern-Simons terms in
odd space-time dimensions, they cannot be written in a (superspace) local
form solely in terms of the field strength.  Finding such superspace non-local 
terms is equivalent to writing the vector multiplet in component fields
and checking supersymmetry invariance ``by hand".  Such terms are known
not to exist up to but not including three derivatives in $N=1$ and $N=2$ 
theories, while it is known that the 2-derivative terms in $N=3$ theories are
in fact superspace Chern-Simons-like terms \cite{r83,gikos85,gio87,gios01}.  
We will report on a 
search for superspace Chern-Simons-like terms in $N=1$ and $N=2$ theories
elsewhere \cite{aabe0307}.  For the remainder of this paper, though, we
ignore the possibility of their existence.

\section{Mixed branch terms}

Terms in the effective action on the $N=2$ mixed branch are simply
terms depending on both the neutral hypermultiplets $\qp$ as well
as the vector multiplets $W$.  As both $\qp$ and $W$ have derivative
dimension $0$, and each is integrated over at least half of the 
Grassmann-odd coordinates in $N=2$ superspace, terms involving either 
superfield have minimum derivative dimension 2.  However, any term 
involving both hyper- and vector multiplets must have dimension greater
than 2 since one is chiral and the other analytic, so they cannot be
integrated over the same half of superspace.  Thus the minimum
dimension term has three derivatives.  In this section we will
construct the dimension three and four terms on the mixed branches
and briefly discuss some of the physics that they describe.

Any 3-derivative term must appear as an integral over
3/4 of the anticommuting coordinates, since the
$\qp$ and $W$ fields have derivative dimension 0.
To be supersymmetric, we must choose the 3/4 of superspace to
be the overlap of the chiral and analytic halves:
\be
S^M_3 = \int\!du\, d^4x\, d^2\thp\, d^2\thbp\, d^2\thm 
\,\,F(\qp, W; u^\pm, \Dpp) + \mbox{c.c}.
\ee
Here $F$ is an arbitrary $\U(1)$-charge +2 function of $W$
and $\qp$ (but not their complex conjugates $\Wb$ and $\qm$).
Since $W$ is $u^\pm$-independent, the $\Dpp$ derivatives
act only on the $\qp$'s.  Just as in the discussion of the
holomorphic Higgs branch term, $S^H_{4a}$, the $\Dmm$ derivatives
do not appear because $\Dmm\qp$ is not analytic.  Also,
if the 2-derivative hypermultiplet kinetic term is free,
then by our previous arguments we need only consider only
the combinations $\qp$, $\Dpp\qp$ and $(\Dpp)^2\qp$ in $F$.

To see in more detail why $S^M_3$ is supersymmetric, recall that up
to total space-time derivatives 
the $d\th^\pm$'s in the Grassmann measure can be replaced by
$D^\mp$'s evaluated at $\th=0$:
\be\label{s3covd}
S^M_3 = \int\!du\, d^4x \left[(\Dm)^2 (\Dbm)^2 (\Dp)^2
\,F((\Dpp)^n\qp, W, u^\pm) \right]_{\th=0} + \mbox{c.c.}.
\ee
This is supersymmetric if it is annihilated by all the supercovariant
derivatives.  Up to total space-time derivatives, it is 
annihilated by $D^\pm$ and $\Dbm$ by antisymmetry (\eg\ $(\Dp)^3=0$).
It is annihilated by the remaining $\Dbp$ since $\Dbp(\Dpp)^n\qp=0$ by 
analyticity and $[\Dbp,\Dpp]=0$, $\Dbp W=0$ by chirality, and $\Dbp u^\pm=0$
by the definition (\ref{Dpmdef}) of $\Dbp$. 

Note that there is a second 3-derivative term given by an integral over
a different three-quarters of superspace,
\be\label{diff34}
S^{'M}_3 = \int\!du\, d^4x\, d^2\thp\, d^2\thbp\, d^2\thbm 
\,\,F'(\qp, \Wb; u^\pm, \Dpp) + \mbox{c.c.} ,
\ee
where $F'$ is now an arbitrary holomorphic function of $\qp$ and
$\Wb$.

These leading 3-derivative supersymmetric terms on mixed branches 
describe a coupling between low energy photons and the hypermultiplet 
and vector multiplet scalars.  To see this, we calculate the bosonic 
part of the action $S^M_3$ in the case where $F$ has no $\Dpp$ dependence
(for simplicity).  We calculate by distributing the covariant
derivatives in (\ref{s3covd}) and using superfield expansions
such as (\ref{qcenexp}) and (\ref{Wphiexp}).
We find that the bosonic part of the action
contains the following two terms 
\bea\label{sm3exp}
S^{M(\mbox{bosonic})}_3 &=& {1\over8} \int\!du\,d^4x 
\left[\ \ F^{IJa}\,
\Dp_\b \Dbm_\a \qp_I \, 
\Db^{-\ad} \Dm_\a \qp_J \,
D^{+\b} D^{-\a} W_a 
\right.\nonumber\\
&&\qquad\qquad \left.\mbox{}+
2 \, F^{IJa}\, 
D^{+\a} \Dp_\a W_a \, 
\Db^{-\ad} D^{-\b} \qp_I \,
\Dbm_\ad \Dm_\b \qp_J 
\right]_{\th=0} +\mbox{c.c.} ,
\eea
where the superscripts on $F$ denote derivatives with respect to
its arguments: $F^I = \del F/\del\qp_I$ and $F^a = \del F/\del W_a$,
where $I$ is an index labeling different hypermultiplets and $a$
labels different vector multiplets.
The last term in (\ref{sm3exp}) contains $(\Dp)^2W$ which is proportional to
the auxiliary $D^{ij}$ field which vanishes by the 2-derivative equations
of motion.  The surviving term gives
\be\label{SM3comp}
S^{M(\mbox{bosonic})}_3 = -
\int\!du\,d^4x\ F^{IJa}(f^+,\phi)\, 
\del_\mu f^+_I\, \del_\nu f^-_J  
\left( F_a^{\mu\nu} + \textstyle{i\over2}
\ep^{\mu\nu\rho\si} F_{a\,\rho\si}\right)
+ \mbox{c.c.}.
\ee
%

We now move on to the 4-derivative terms on the mixed branch.
To make 4-derivative terms given as integrals over 3/4 of
superspace as in $S^M_3$, we require derivative dimension 1
combinations covariant derivatives of $\qp$ and $W$ annihilated 
by $\Dbp$.  There are five\footnote{Two other scalar combinations,
$(\Dbp)^2\Wb$ and $(\Dbp)^2\qm$, also give rise to supersymmetric
4-derivative terms, but they are just special cases of the
non-holomorphic 4-derivative term $S^M_{4c}$ given in (\ref{sm4c}).}
such combinations:  $\del_\mu\qp$,
$\del_\mu W$, $\Dp W \cdot \si^\mu\cdot \Dbm \qp$, $\Dp W\cdot\Dp W$, and
$\Dbm\qp\cdot\Dbm\qp$.  The first three are not Lorentz invariant, so
can be dropped.  The second two then give rise to the following
4-derivative terms:
\bea\label{sm4ab}
S^M_{4a} &=& \int\!du\, d^4x\, d^2\thp\, d^2\thbp\, d^2\thm 
\,\,\Dp W_a\cdot\Dp W_b\ G^{ab}(\qp, W; u^\pm, \Dpp) + \mbox{c.c.}\\
S^M_{4b} &=& \int\!du\, d^4x\, d^2\thp\, d^2\thbp\, d^2\thm 
\,\,\Dbm(\Dpp)^n\qp_I\cdot \Dbm(\Dpp)^m\qp_J \  
G_{nm}^{IJ}(\qp, W; u^\pm, \Dpp) 
+ \mbox{c.c.} \nonumber
\eea
There are also versions of each of these terms integrated over a different
three-quarters of superspace, as in (\ref{diff34}).
Finally, there is also a non-holomorphic 4-derivative term given by an
integral over all of superspace:
\be\label{sm4c}
S^M_{4c}=\int\! du\,d^4x\,d^8\th \,\,H(\qp,\qm,W,\Wb; u^\pm, D^{\pm\pm}).
\ee

The expressions (\ref{sm4ab}) and (\ref{sm4c}) are non-local on the
auxiliary harmonic $u$-sphere, since an arbitrary number of
$D^{\pm\pm}$ derivatives appear.  But, just as was discussed in
section 1 and in section 4 following eqn.~(\ref{4der}), we can
eliminate this non-locality by consistently using the auxiliary
field equations of motion following from 2-derivative terms.  
In the case where the 2-derivative hypermultiplet kinetic term
is free, we can limit the appearance of $D^{\pm\pm}$ to the
two combinations $\Dpp\qp$ and $(\Dpp)^2\qp$ in the holomorphic
terms $S^M_{4a,b}$, and to the ten combinations $(\Dpp)^n\qp$,
$(\Dmm)^m\qp$ for $n=1,2$ and $m=1,2,3$, and their complex conjugates
in $S^M_{4c}$.

Which of $S^M_{4a,b,c}$ can give rise to a WZ term?
Following the discussion given in the previous
sections, we saw that WZ terms are proportional to
$\ep^{\mu\nu\rho\si} \propto \mbox{tr} (\si^\mu \bar{\si}^\nu 
\si^{\rho} \bar{\si}^\si)$,
and must therefore contain all four types ($\dpp$, 
$\dpm$, $\dmm$, and $\dmp$) of derivatives acting on the scalar fields. 
Three independent combinations of those derivatives occurred in the
hypermultiplet component expansion (\ref{qcenexp}) and the fourth
occurs in the vector multiplet component expansion (\ref{Wphiexp}). 
Therefore $S^M_{4a,b,c}$ could each contain a WZ term.  

Recall the discussion of section 4 where it was shown that for
WZ terms only $q^\pm$ and $D^{\pm\pm}q^\mp \sim \qt^\pm$ can
contribute.  This implies that of $S^M_{4a,b,c}$ only terms of
the form
\bea
S^{M\mbox{\tiny (WZ)}}_{4a} &=& \int\!du\, d^4x\, d^2\thp\, d^2\thbp\, d^2\thm 
\,\,\Dp W_a\cdot\Dp W_b\ G^{ab}(\qp, W; u^\pm) + \mbox{c.c.},
\nonumber\\
S^{M\mbox{\tiny (WZ)}}_{4b} &=& \int\!du\, d^4x\, d^2\thp\, d^2\thbp\, d^2\thm 
\,\,\Dbm\qp_I\cdot \Dbm\qp_J \  
G_{00}^{IJ}(\qp, W; u^\pm) + \mbox{c.c.},
\nonumber\\
S^{M\mbox{\tiny (WZ)}}_{4c} &=& \int\! du\,d^4x\,d^8\th \,\,
H(\qp,\qtp,\qm,\qtm,W,\Wb; u^\pm),
\eea
need be considered.

It is not hard to see that $S^{M\mbox{\tiny (WZ)}}_{4a}$ cannot contribute a
WZ term.  From (\ref{Wphiexp}) the bosonic expansion of $\Dp W_a$ is
proportional to at least one factor of $\thbp$.  Thus $(\Dbm)^2$ from the
superspace measure must hit the $(\Dp W)^2$ factor, giving 
$\Dbm_\ad \Dp_\a W \Db^{-\ad} D^{+\a} W \propto \del_\mu W \del^\mu W$.
Since the space-time derivatives are contracted, this cannot give rise
to a WZ term.

A similar argument shows that $S^{M\mbox{\tiny (WZ)}}_{4b}$ does not contribute
a WZ term either.  From (\ref{qcenexp}), we have $\Dbm_\ad\qp = -2i
(\thp\dsl)_\ad f^- + {\cal O}(\del^2f)$, implying that $(\Dm)^2$ from
the measure must hit the $(\Dbm\qp)^2$ factor to absorb the $\thp$'s.
Since $\Dm_\a \Dbm_\ad \qp = 2i \dsl_{\a\ad} f^- + {\cal O}(\del^2f)$, we
have $\Dm_\a \Dbm_\ad\qp D^{-\a} \Db^{-\ad} \qp \propto \del_\mu f^- 
\del^\mu f^-$, which has contracted space-time derivatives and so
cannot contribute a WZ term.

The scalar component expansion of $S^{M\mbox{\tiny (WZ)}}_{4c}$ can be performed
in a similar way.  Since both $q^\pm$ and $\qt^\pm$ appear in $H$, we use
the compact notation (introduced in section 4.2 above) where the indices
$I,J,K,L$ run over the $q$'s as well as the $\qt$'s.  Thus 
\be
\qp_L  \equiv (\qp_\ell, -\qtp_\ell),
\qquad\qquad
\qm_\Lbar  \equiv (\qm_\ell, -\qtm_\ell).
\ee
For the scalar components we define 
\be
f^\pm_L \equiv (f^\pm_\ell, -\fb^\pm_\ell), 
\qquad\mbox{and}\qquad 
\fb^\pm_L \equiv (-\fb^\pm_\ell, -f^\pm_\ell),
\ee
so that $\qp_L|_{\th=0} = \fp_L$ and $\qm_\Lbar|_{\th=0}=\fbm_\Lbar$
and
\be\label{fpmcc}
\bar{f^\pm_L} = \pm\fb^\mp_\Lbar ,
\ee
which follows from (\ref{ucc}) and (\ref{fpmdefs}).

In order to get
the right Lorentz structure, the eight supercovariant derivatives from
the measure must
act in $\Db D$ pairs on four different fields.  If we choose to act with the
$D$'s first, then none of those four fields will be a $\Wb$ since they
are annihilated by $\Db$'s.  Also, by analyticity, $\Dp$ annihilates
$\qp$ and $\Dm$ annihilates $\qm$.  Finally, to get the epsilon tensor, we
need a trace of four sigma matrices, so that the spinor indices must be
contracted to give a single trace.  All these constraints mean that
there are only four ways of distributing the covariant derivatives,
giving
\bea\label{WZexpansion}
S^{M\mbox{\tiny (WZ)}}_{4c} &=& -{1\over16}\int\! du\,d^4x \left[
H^{\Ibar\Jbar KL}
\,\ \Db^{+\ad} D^{+\a} \qm_\Ibar
\,\ \Dbm_\bd   \Dp_\a  \qm_\Jbar  
\,\ \,\Db^{-\bd} D^{-\b} \qp_K
\,\ \Dbp_\ad \Dm_\b \qp_L
\right.\nonumber\\
&&\qquad\qquad\quad \,\left.\mbox{}
+H^{\Ibar\Jbar Ka}
\,\ \Db^{+\ad} D^{+\a} \qm_\Ibar
\,\ \Dbm_\bd   \Dp_\a  \qm_\Jbar  
\,\ \,\Db^{-\bd} D^{-\b} \qp_K
\,\ \Dbp_\ad \Dm_\b W_a
\right.\nonumber\\
&&\qquad\qquad\quad \,\left.\mbox{}
+H^{\Ibar aKL}
\,\ \Db^{+\ad} D^{+\a} \qm_\Ibar
\,\ \Dbm_\bd   \Dp_\a  W_a
\,\ \Db^{-\bd} D^{-\b} \qp_K
\,\ \Dbp_\ad \Dm_\b \qp_L
\right.\nonumber\\
&&\qquad\qquad\quad \,\left.\mbox{}
+H^{\Ibar aKb}
\,\ \Db^{+\ad} D^{+\a} \qm_\Ibar
\,\ \Dbm_\bd   \Dp_\a  W_a
\,\ \Db^{-\bd} D^{-\b} \qp_K
\,\ \Dbp_\ad \Dm_\b W_b
\ \right]\nonumber\\
&=& 2 \ep^{\mu\nu\rho\si} \int\! du\,d^4x  \left[
\ \ \,i\ \, H^{\Ibar\Jbar KL}
\ \del_\mu \fbp_\Ibar 
\ \del_\nu \fbm_\Jbar
\ \del_\rho \fm_K
\ \del_\si \fp_L
\right. \nonumber\\
&&\qquad\qquad\qquad\ \mbox{} 
- \sqrt2\, H^{\Ibar\Jbar Ka}\,
\ \del_\mu \fbp_\Ibar
\ \del_\nu \fbm_\Jbar
\ \del_\rho \fm_K
\ \del_\si \phi_a \nonumber\\
&&\qquad\qquad\qquad\ \mbox{} 
- \sqrt2\, H^{\Ibar aKL}\,
\ \del_\mu \fbp_\Ibar
\ \del_\nu \phi_a
\ \,\del_\rho \fm_K
\ \del_\si \fp_L \nonumber\\
&&\qquad\qquad\qquad\ 
\left. \mbox{}-\,2 i\ H^{\Ibar aKb}\,\ 
\ \del_\mu \fbp_\Ibar
\ \del_\nu \phi_a
\ \,\del_\rho \fm_K
\ \del_\si \phi_b 
\ \right],
\eea
where the superscripts on $H$ denote derivatives with respect to its
arguments, as before.  In the second equality we used that
\bea
\Db^\pm_\ad \Dm_\a \qp_L = + 2i \dsl_{\a\ad} f^\pm_L + {\cal O}(\del^2f),
\ \ \, &\qquad&
\Db^\pm_\ad \Dp_\a \qm_\Lbar = - 2i \dsl_{\a\ad} \fb^\pm_\Lbar + {\cal O}(\del^2f),
\nonumber\\
\Db^\pm_\ad D^\mp_\a W_a = \mp 2\sqrt2\,\dsl_{\a\ad}\phi_a + {\cal O}(\del^2\phi),
&\qquad&
\Db^\pm_\ad D^\pm_\a W_a = {\cal O}(\del^2\phi),
\eea
which follow from the scalar field expansions (\ref{qcenexp}) and (\ref{Wphiexp})
of $q^\pm$ and $W$ to first order in derivatives.  We also used the sigma
matrix identity tr$(\si^\mu \bar\si^\nu \si^\rho \bar\si^\si) = -2i 
\ep^{\mu\nu\rho\si} + 2\eta^{\mu\nu}\eta^{\rho\si}  - 2\eta^{\mu\rho}\eta^{\nu\si} 
+ 2\eta^{\mu\si}\eta^{\nu\rho}$, and kept only the $\ep^{\mu\nu\rho\si}$ piece.  

The expression (\ref{WZexpansion}) can be further simplified.  The fourth 
term in the second equality cancels by the antisymmetry on $\nu$, $\si$ and 
the symmetry on $a$, $b$.  Furthermore, by (\ref{fpmcc}) and the reality of 
$H$, it follows that the first term in the second equality in (\ref{WZexpansion})
is purely imaginary.  Since the original action was real, this imaginary
term must be part of a total derivative introduced when we replaced the
$d\th$ integrations by covariant derivatives.  Indeed, it is not too hard
to see that the first term plus the imaginary part of the second and
third terms are a total derivative, and can therefore be dropped.  Thus
the Wess Zumino term is the real parts of the second and third terms in
(\ref{WZexpansion}), which can be rewritten
\bea
S^{M\mbox{\tiny (WZ)}}_{4c} &=&
-\sqrt2\, \ep^{\mu\nu\rho\si} \!\!\int\! du\,d^4x  
\ \del_\mu \fbp_\Ibar
\, \del_\rho \fm_K 
\Bigl( \del_\nu \fbm_\Jbar\, \del^\Jbar 
+  \del_\nu \fp_J\, \del^J \Bigr)
\Bigl(\del_\si \phi_a\, \del^a 
+ \del_\si \bar\phi_{\bar a}\, \del^{\bar a} \Bigr) 
H^{\Ibar K}
\nonumber\\
&=& 
2\sqrt2\, \ep^{\mu\nu\rho\si} \!\!\int\! du\,d^4x  
\  \del_\mu  \fbp_\Ibar
\, \del_\nu  \fbm_\Jbar
\, \del_\rho \fm_K 
\, \del_\si  \fp_L
\, H^{\Ibar\Jbar KL} ,
\eea
where in the second line we integrated $\del_\si$ by parts.
Since this is not a total derivative (as long as $H$ depends on $W$ and
$\Wb$), we have
shown that $S^M_{4c}$ is the $N=2$ supersymmetric completion
of the Wess-Zumino term.

\acknowledgments It is a pleasure to thank S. Das, R. Plesser, and A. Shapere 
for helpful comments and discussions, and K. Meyers for sending us reference
\cite{gios01}.  This work was supported in part by DOE grant
DOE-FG02-84ER-40153.  G.A.B. was supported in part by a University of 
Cincinnati URC grant.

\appendix
\section{Grassmann locality from space-time locality and supersymmetry}

In this appendix we will present an argument showing that the combination
of space-time locality with supersymmetry implies that expressions involving
unconstrained superfields are necessarily local in the Grassmann-odd coordinates.
Though this seems like a fundamental property of superspaces, we do not know
of a reference for this argument.

To keep the notation simple, we give the argument in $N=2$ supersymmetric
quantum mechanics; the generalization to any superspace is straight forward.
The superspace then consists of a space-time coordinate $x$ and two 
Grassmann-odd coordinates $\th$ and $\thb$.  Denote derivatives by
\be
d={\del\over\del x},\qquad
\del = {\del\over\del\th},\qquad \mbox{and} \quad
\delb = {\del\over\del\thb} .
\ee
The supercharges
\be\label{Q1def}
Q= \del - \thb d,
\qquad\mbox{and}\qquad
\Qb = \delb - \th d ,
\ee
generate the supersymmetry algebra
\be
\{Q,\Qb\}=-2d ,
\ee
and the supercovariant derivatives
\be\label{D1def}
D= \del + \thb d,
\qquad\mbox{and}\qquad
\Db = \delb + \th d ,
\ee
anticommute with the supercharges.  The supercharges generate translations 
and supertranslations of general superfields $\phi(x,\th,\thb)$ according to
\be
\de\phi = (\a d + \ep Q + \bar\ep\Qb)\phi ,
\ee
where $\a$, $\ep$, and $\bar\ep$ are arbitrary constants.

The general (non-local) term in an action for unconstrained
superfields can be written
\be\label{nonlocalterm}
S = \int\!dx_1d\th_1d\thb_1\cdots dx_nd\th_nd\thb_n\
{\cal K}(x_1,\th_1,\thb_1;\ldots;x_n,\th_n,\thb_n)\ 
\phi_1(x_1,\th_1,\thb_1)\cdots\phi_n(x_n,\th_n,\thb_n) ,
\ee
where $\cal K$ is an arbitrary kernel, and the $\phi_i$'s stand
for arbitrary unconstrained superfields and their derivatives.  
The general action will be the sum of many such terms.  If
the $\phi_i$'s are unconstrained superfields, then 
super-Poincar\'e invariance implies $\de S=0$ for each term
individually, since each term has a different functional dependence
on the superfields.  If, however, the superfields were constrained, so
that there were functional relations among them, then we could only
demand super-Poincar\'e invariance of the whole sum, not necessarily
for each individual term, and the following argument would not work.

So, for unconstrained superfields we have
\bea\label{a7}
0 &=& \de S = \int\!dx_1\cdots d\thb_n\
{\cal K}(x_1,\ldots,\thb_n)\ 
\de (\phi_1\cdots\phi_n) 
\nonumber\\
&=& \int\!dx_1\cdots d\thb_n\
{\cal K}(x_1,\ldots,\thb_n)\ 
\sum_{i=1}^n (\a d_i+\ep Q_i+\bar\ep \Qb_i)
\ (\phi_1\cdots\phi_n)
\nonumber\\
&=& -\int\!dx_1\cdots d\thb_n\
(\phi_1\cdots\phi_n)\ 
\sum_{i=1}^n (\a d_i+\ep Q_i+\bar\ep \Qb_i)
\ {\cal K}(x_1,\ldots,\thb_n) ,
\eea
where $Q_i$ refers to the derivative operator (\ref{Q1def}) acting on
$\{x_i,\th_i\thb_i\}$, and where in the last line we performed and
integration by parts.  Because $\a$, $\ep$, and $\bar\ep$ are independent
arbitrary constants, and because the $\phi_i$'s are unconstrained, (\ref{a7}) 
implies that the kernel must be separately annihilated by the sums of
the super-Poincar\'e generators:
\be\label{susycons}
0 
= \left(\sum_{i=1}^n d_i \right){\cal K} 
= \left(\sum_{i=1}^n Q_i \right){\cal K} 
= \left(\sum_{i=1}^n \Qb_i \right){\cal K}  . 
\ee
The general solution to these supersymmetry equations is that 
$\cal K$ depends only on the combinations
\be
{\cal K} = {\cal K}(\, x_i - x_j + \th_j\thb_i- \th_i\thb_j 
\, , \, \th_i - \th_j \, , \, \thb_i - \thb_j\,).
\ee

Space-time locality implies that $\cal K$ must have support
only for $x_i=x_j$, \ie,
\be\label{Kexp}
{\cal K} = \til{\cal K}( \th_i - \th_j \, , \, \thb_i - \thb_j )\ 
\prod_{i>j}\de(x_i - x_j + \th_j\thb_i- \th_i\thb_j ) .
\ee
Now, by the anticommuting nature of the $\th$'s, 
\be
\th_i-\th_j=\de(\th_i-\th_j),
\ee
and similarly for the $\thb$'s.  So any non-trivial $\til{\cal K}$
factor in (\ref{Kexp}) just enforces (some) locality in the Grassmann-odd
coordinates.  Since we are trying to show just such locality, we need
only concentrate on the delta-function factor in (\ref{Kexp}).

It is sufficient to focus on any pair of $(i,j)$.  Denote these two
superspace points by $(x,\th,\thb)$ and $(x',\th',\thb')$, and rename the
two superfields $\phi_i,\phi_j$ to $\phi$ and $\psi$, respectively.
Thus we are interested in the expression
\be
I = \int\!dxd\th d\thb dx'd\th'd\thb'\,\de(x-x'+\th'\thb-\th\thb') \,
\phi(x,\th,\thb)\,\psi(x',\th',\thb') 
\ee
for general superfields $\phi$ and $\psi$.  We will show that $I=J$ where
$J$ is 
\be
J = {1\over2}\int\!dxd\th d\thb\,\phi(x,\th,\thb) \,[D,\Db]\, \psi(x,\th,\thb) ,
\ee
and is thus local in the Grassmann-odd coordinates.  Because $\phi[D,\Db]\psi$
is itself another superfield, this argument can then be repeated with other
pairs of superspace coordinates until the whole expression (\ref{nonlocalterm})
is written as a single integral over superspace.

To show this is a straight forward computation:
\bea\label{asfc}
I &=& \int\!dxd\th d\thb dx'd\th'd\thb'\,
\phi(x,\th,\thb)
\,\de(x-x'+\th'\thb-\th\thb') 
\,\psi(x',\th',\thb') 
\nonumber\\
&=& \int\!dxd\th d\thb dx'd\th'd\thb'\,
\phi(x,\th,\thb)
\de(x-x')\left\{ 1+(\th'\thb-\th\thb')d'
+\th\thb\th'\thb'(d')^2 \right\}
\,\psi(x',\th',\thb') 
\nonumber\\
&=& \int\!dx \del\delb\del'\delb'\left[
\phi(x,\th,\thb)
\left\{ 1+(\th'\thb-\th\thb')d
+\th\thb\th'\thb' d^2 \right\}
\,\psi(x,\th',\thb') \right]_{\th=\thb=\th'=\thb'=0}
\nonumber\\
&=& \int\!dx \left[
\del\delb\phi \, \del'\delb'\psi
+ \del\phi \, \delb' d\psi
+ \delb\phi \, \del' d\psi
+ \phi d^2 \psi 
\right]_{\th=\thb=\th'=\thb'=0}
\nonumber\\
&=& \int\!dx \left[
\del\delb\phi\, \del\delb\psi
+ \del\phi \,  \delb d \psi
+ \delb\phi \, \del d \psi
+ \phi d^2 \psi 
\right]_{\th=\thb=0},
\eea
where in the second line we Taylor expanded the delta function
and integrated by parts in $x'$; in the third line we performed the $x'$
integration and replaced the $\th$ integrations by derivatives evaluated
at zero; in the fourth line we expanded the derivatives keeping only
terms that survive when the $\th$'s are set to zero; and in the last line
we replaced $\th'$ and $\thb'$ with $\th$ and $\thb$ since they are
all set to zero anyway.  On the other hand,
\bea\label{Jasfc}
J &=& \int\!dxd\th d\thb
\,\phi(x,\th,\thb) 
\,\textstyle{1\over2}[D,\Db]
\,\psi(x,\th,\thb) 
\nonumber\\
&=& \int\!dxd\th d\thb
\,\phi 
\,(\del\delb-\th\del d+\thb\delb d-\th\thb d^2)
\,\psi
\nonumber\\
&=& \int\!dx\,\del\delb \left[
\phi \,\del\delb\psi
- \phi \,\th\del d\psi
+ \phi \,\thb\delb d\psi
- \phi \,\th\thb d^2\psi 
\right]_{\th=\thb=0}
\nonumber\\
&=& \int\!dx \left[
\del\delb\phi \,\del\delb\psi
+ \delb\phi \,\del d\psi
+ \del\phi \,\delb d\psi
+ \phi \,d^2\psi 
\right]_{\th=\thb=0} ,
\eea
where in the second line we expanded $[D,\Db]$ using (\ref{D1def});
in the third line we replaced the $\th$ integrations by derivatives 
evaluated at zero; and in the last line we expanded the derivatives 
keeping only terms that survive when the $\th$'s are set to zero.
Thus $I=J$.

If some of the superfields are unconstrained functions over only a 
subspace of the full superspace, the same type of argument applies.  
For example, say $\psi(x',\th',\thb')$ is chiral, so that $\Db'\psi=0$.
Then $\psi=\psi(x_C',\th')$ where $x_C'=x'+\th'\thb'$.  A typical
term in the action will have the general form
\be
S = \int\!dxd\th d\thb\,dx_C'd\th'
\ {\cal K}(x,\th,\thb;x_C',\th')
\ \phi(x,\th,\thb)\, \psi(x_C',\th'),
\ee
and supersymmetry will imply the same constraints (\ref{susycons}) on 
$\cal K$ as before.  The solution is different, though, since $\cal K$
only depends on $\thb'$ through $x_C'$, giving that $\cal K$ is a function
only of the combinations $x-x_C'+2\th'\thb-\th\thb$ and $\th=\th'$.
Space-time locality then implies that we are interested in the expression
\bea
I &\equiv& \int\!dxd\th d\thb\,dx_C'd\th'
\ \phi(x,\th,\thb)
\ \de(x-x_C'+2\th'\thb-\th\thb)
\ \psi(x_C',\th')
\nonumber\\
&=& \int\!dxd\th d\thb\,dx_C'd\th'
\ \phi(x,\th,\thb)
\ \de(x-x_C') 
\left\{1+(2\th'\thb-\th\thb)d'_C\right\}
\ \psi(x_C',\th')
\nonumber\\
&=& \int\!dx\del\delb\del' \left[
\phi(x,\th,\thb)
\left\{1+(2\th'\thb-\th\thb)d\right\}
\psi(x,\th') \right]_{\th=\thb=\th'=0}
\nonumber\\
&=& \int\!dx \left[
\del\delb\phi \, \del'\psi
+ 2 \del\phi \, d\psi
+ \phi \, \del'd\psi 
\right]_{\th=\thb=\th'=0}
\nonumber\\
&=& \int\!dx \left[
\del\delb\phi \, \del\psi
+ 2 \del\phi \, d\psi
+ \phi \, \del d\psi 
\right]_{\th=\thb=0} ,
\eea
where we have followed the same steps as in (\ref{asfc}).
On the other hand, 
\bea
J &\equiv& \int\!dx\th d\thb
\ \phi(x,\th,\thb)\ D\,\psi(x_C,\th)
\nonumber\\
&=& \int\!dx\th d\thb
\ \phi\ (\del+\thb d)\psi(x+\th\thb,\th)
\nonumber\\
&=& \int\!dx\del\delb \left[
\phi\del\psi+2\phi\thb d\psi
\right]_{\th=\thb=0}
\nonumber\\
&=& \int\!dx \left[
\del\delb\phi\, \del\psi
- \phi\, \del d\psi
+ 2\del\phi\, d\psi
+ 2\phi\, \del d\psi
\right]_{\th=\thb=0} ,
\eea
where we followed the same steps as in (\ref{Jasfc}), though it 
should be pointed out that inside the square brackets $\del$ and
$d$ refer to derivatives of $\psi$ with respect to its 
arguments---\ie\ partial derivatives and not total derivatives.
Thus $I=J$ and we have shown that general terms involving both 
chiral and non-chiral unconstrained superfields are given by
local superspace expressions.  Similar arguments take care of the
other cases (chiral-chiral and chiral-antichiral).

\end{document}